\documentclass[11pt,fleqn]{article}
\usepackage{amsfonts, epsfig, amssymb, amsmath}
\usepackage[normalem]{ulem}
\usepackage{color}
\newcommand{\ol}{\setlength{\itemsep}{0pt.}\begin{enumerate}}
\newcommand{\eol}{\end{enumerate}\setlength{\itemsep}{-\parsep}}
\newcommand{\ignore}[1]{}
\title{One more proof of the first linear programming bound for binary codes and two conjectures}
\author{Alex Samorodnitsky}

\begin{document}
\date{}
\maketitle


\newtheorem{THEOREM}{Theorem}[section]
\newenvironment{theorem}{\begin{THEOREM} \hspace{-.85em} {\bf :}
}%
                        {\end{THEOREM}}
\newtheorem{LEMMA}[THEOREM]{Lemma}
\newenvironment{lemma}{\begin{LEMMA} \hspace{-.85em} {\bf :} }%
                      {\end{LEMMA}}
\newtheorem{COROLLARY}[THEOREM]{Corollary}
\newenvironment{corollary}{\begin{COROLLARY} \hspace{-.85em} {\bf
:} }%
                          {\end{COROLLARY}}
\newtheorem{PROPOSITION}[THEOREM]{Proposition}
\newenvironment{proposition}{\begin{PROPOSITION} \hspace{-.85em}
{\bf :} }%
                            {\end{PROPOSITION}}
\newtheorem{DEFINITION}[THEOREM]{Definition}
\newenvironment{definition}{\begin{DEFINITION} \hspace{-.85em} {\bf
:} \rm}%
                            {\end{DEFINITION}}
\newtheorem{EXAMPLE}[THEOREM]{Example}
\newenvironment{example}{\begin{EXAMPLE} \hspace{-.85em} {\bf :}
\rm}%
                            {\end{EXAMPLE}}
\newtheorem{CONJECTURE}[THEOREM]{Conjecture}
\newenvironment{conjecture}{\begin{CONJECTURE} \hspace{-.85em}
{\bf :} \rm}%
                            {\end{CONJECTURE}}
\newtheorem{MAINCONJECTURE}[THEOREM]{Main Conjecture}
\newenvironment{mainconjecture}{\begin{MAINCONJECTURE} \hspace{-.85em}
{\bf :} \rm}%
                            {\end{MAINCONJECTURE}}
\newtheorem{PROBLEM}[THEOREM]{Problem}
\newenvironment{problem}{\begin{PROBLEM} \hspace{-.85em} {\bf :}
\rm}%
                            {\end{PROBLEM}}
\newtheorem{QUESTION}[THEOREM]{Question}
\newenvironment{question}{\begin{QUESTION} \hspace{-.85em} {\bf :}
\rm}%
                            {\end{QUESTION}}
\newtheorem{REMARK}[THEOREM]{Remark}
\newenvironment{remark}{\begin{REMARK} \hspace{-.85em} {\bf :}
\rm}%
                            {\end{REMARK}}

\newcommand{\thm}{\begin{theorem}}
\newcommand{\lem}{\begin{lemma}}
\newcommand{\pro}{\begin{proposition}}
\newcommand{\dfn}{\begin{definition}}
\newcommand{\rem}{\begin{remark}}
\newcommand{\xam}{\begin{example}}
\newcommand{\cnj}{\begin{conjecture}}
\newcommand{\mcnj}{\begin{mainconjecture}}
\newcommand{\prb}{\begin{problem}}
\newcommand{\que}{\begin{question}}
\newcommand{\cor}{\begin{corollary}}
\newcommand{\prf}{\noindent{\bf Proof:} }
\newcommand{\ethm}{\end{theorem}}
\newcommand{\elem}{\end{lemma}}
\newcommand{\epro}{\end{proposition}}
\newcommand{\edfn}{\bbox\end{definition}}
\newcommand{\erem}{\bbox\end{remark}}
\newcommand{\exam}{\bbox\end{example}}
\newcommand{\ecnj}{\bbox\end{conjecture}}
\newcommand{\emcnj}{\bbox\end{mainconjecture}}
\newcommand{\eprb}{\bbox\end{problem}}
\newcommand{\eque}{\bbox\end{question}}
\newcommand{\ecor}{\end{corollary}}
\newcommand{\eprf}{\bbox}
\newcommand{\beqn}{\begin{equation}}
\newcommand{\eeqn}{\end{equation}}
\newcommand{\wbox}{\mbox{$\sqcap$\llap{$\sqcup$}}}
\newcommand{\bbox}{\vrule height7pt width4pt depth1pt}
\newcommand{\qed}{\bbox}

\def\H{\{0,1\}^n}

\def\S{S(n,w)}

\def\g{g_{\ast}}
\def\xop{x_{\ast}}
\def\y{y_{\ast}}
\def\z{z_{\ast}}

\def\f{\tilde f}

\def\n{\lfloor \frac n2 \rfloor}

\def \E{\mathop{{}\mathbb E}}
\def \R{\mathbb R}
\def \N{\mathbb N}
\def \Z{\mathbb Z}
\def \F{\mathbb F}
\def \T{\mathbb T}

\ignore{\def \x{\textcolor{red}{x}}
\def \r{\textcolor{red}{r}}
\def \Rc{\textcolor{red}{R}}
}

\def \noi{{\noindent}}

\def\myblt{\noi --\, }

\def \iff{~~~~\Leftrightarrow~~~~}

\def \queq {\quad = \quad}

\def\<{\left<}
\def\>{\right>}
\def \({\left(}
\def \){\right)}

\def \e{\epsilon}
\def \l{\lambda}

\def\Tp{Tchebyshef polynomial}
\def\Tps{TchebysDeto be the maximafine $A(n,d)$ l size of a code with distance $d$hef polynomials}
\newcommand{\rarrow}{\rightarrow}

\newcommand{\larrow}{\leftarrow}

\overfullrule=0pt
\def\setof#1{\lbrace #1 \rbrace}

\begin{abstract}

We give one more proof of the first linear programming bound for binary codes, following the line of work initiated by Friedman and Tillich \cite{ft}. The new argument is somewhat similar to the one given in \cite{ns1}, but we believe it to be both simpler and more intuitive. Moreover, it provides the following 'geometric' explanation for the bound. A binary code with minimal distance $\delta n$ is small because the projections of the characteristic functions of its elements on the subspace spanned by the Walsh-Fourier characters of weight up to $\(\frac 12 - \sqrt{\delta(1-\delta)}\) \cdot n$ are essentially independent. Hence the cardinality of the code is bounded by the dimension of the subspace.

We present two conjectures, suggested by the new proof, one for linear and one for general binary codes which, if true, would lead to an improvement of the first linear programming bound. The conjecture for linear codes is related to and is influenced by conjectures of H{\aa}stad and of Kalai and Linial. We verify the conjectures for the (simple) cases of random linear codes and general random codes.
\end{abstract}

\section{Introduction}

\noi A binary error-correcting code $C$ of length $n$ and minimal distance $d$ is a subset of the Hamming cube $\H$ in which the distance between any two distinct points is at least $d$. Let $A(n, d)$ be the maximal size of such a code.
In this paper we are interested in the case in which the distance $d$ is linear in the length $n$ of the code, and we let $n$ go to infinity.
In this case $A(n, d)$ is known (see e.g., \cite{lev-chapter}) to grow exponentially in $n$, and we consider the quantity
\[
R(\delta) ~=~ \limsup_{n \rarrow \infty} \frac 1n \log_2 A\(n,\lfloor \delta n \rfloor\),
\]
also known as the {\it asymptotic maximal rate} of the code with relative distance $\delta$, for $0 \le \delta \le \frac12$.

\noi The best known upper bounds on $R(\delta)$ were obtained in \cite{mrrw} using the linear programming relaxation, constructed in \cite{dels}, of the combinatorial problem of bounding $A(n,d)$. While the precise value of the linear program of \cite{dels} is still unknown, there is a convincing numerical evidence \cite{bj} that on the exponential scale the bounds of \cite{mrrw} are the best possible to derive from this program. It follows that in order to improve these bounds we need either to augment the linear program of \cite{dels}, or to look for a different way to prove the bounds of \cite{mrrw}. The first approach was adopted by \cite{schrijver} (see also e.g., \cite{bgsv}), who suggested a positive semidefinite relaxation of the problem to bound $A(n,d)$, augmenting the linear programming relaxation of \cite{dels} by studying the geometry of a code in more detail (see also the discussion before Proposition~\ref{pro:general-good} below). The second approach was taken by \cite{ft}, where the {\it first linear programming bound} for linear binary codes was proved by a different and a more direct argument. Given a linear code $C$, that is, a linear subspace of the Hamming cube, \cite{ft} proved comparison theorems (adapting ideas from Riemannian geometry to the discrete setting) between two metric spaces defined by the two Cayley graphs: The Hamming cube $\H$ and the Cayley graph of the quotient $\H / C^{\perp}$ with respect to the set of generators given by the standard basis of $\H$. The key observations were that the metric balls in $\H$ grow faster than their counterparts in $\H / C^{\perp}$, while their {\it  eigenvalues} (the eigenvalue of a set is the maximal eigenvalue of the adjacency matrix of the graph restricted to this set) are bounded from above by these of their counterparts.

\noi Following \cite{ft}, where the importance of working with Hamming balls and their eigenvalues was established, the expediency of working with the maximal {\it eigenfunctions} of Hamming balls was observed in \cite{ns, ns1}. It was, in effect, shown that, given any nonnegative function $f$ on $\H$ with a small support, such that the adjacency matrix of the Hamming cube acts on $f$ by multiplying it pointwise by a large factor, one can obtain an upper bound on the cardinality of error-correcting codes, whose applicability will depend on the cardinality of the support of $f$ and on the size of the multiplying factor. Using the maximal eigenfunctions of Hamming balls of different radii, with their corresponding parameters, led to a simple proof of the first linear programming bound for linear codes, which was then extended to prove the bound for general binary codes as well. One appealing feature of the argument for linear codes was that it established the following 'covering' explanation for the first linear programming bound (stated explicitly in \cite{ns1} and contained implicitly in \cite{ft}). A linear code $C$ with minimal distance $d$ is small, because its dual $C^{\perp}$ is large, in the following sense: A union of Hamming balls of radius $r = r(d)$ centered at the points of $C^{\perp}$ covers almost the whole space. This implies that, up to negligible errors, $|C| \le |B|$, where $B$ is a ball of radius $r$. Unfortunately, the extension of the argument to general binary codes seemed to allow no such natural geometric interpretation.

\noi In this paper we give another proof of the first linear programming bound for general binary codes. This proof is somewhat similar to the one given in \cite{ns1}, but we believe it to be both simpler and more satisfactory, in that it provides a 'geometric' explanation for the bound. We show that a binary code with minimal distance $\delta n$ is small because the projections of the characteristic functions of its elements on the subspace spanned by the Walsh-Fourier characters of weight up to $\(\frac 12 - \sqrt{\delta(1-\delta)}\) \cdot n$ are essentially independent. Hence the cardinality of the code is essentially bounded by the dimension of the subspace.
Let $0 \le r \le n$, and let $\Lambda_r$ be the orthogonal projection on the span of the Walsh-Fourier characters of weight at most $r$ (see Section~\ref{subsec:backg} for background and definitions of relevant notions). For $x \in \H$, let $\delta_x$ be the characteristic function of the point $x$. Let $\<v_1,...,v_N\>$ denote the linear span of the vectors $v_1,...,v_N$. We prove the following claim.

\thm
\label{thm:JPL1}
Let $0 < \delta < \frac12$. There exists a function $r = r_{\delta} : \N \rarrow \N$, with $r(n) = \(\frac 12 - \sqrt{\delta(1-\delta)}\) \cdot n + o(n)$, such that for any code $C$ of length $n$ and minimal distance $d = \lfloor \delta n\rfloor$ holds
\[
\dim\bigg(\Big<\big\{\Lambda_{r(n)} \delta_x\big\}_{x \in C}\Big>\bigg) ~\ge~ \frac{1}{2d} \cdot |C|.
\]
\ethm

\noi Let us make several comments about this result.

\myblt It follows that $|C| \le 2d \cdot \sum_{k=0}^{r(n)} {n \choose k}$. By the known exponential estimates for binomial coefficients (see (\ref{binom}) below) this implies the first linear programming bound on the asymptotic rate function:
$R(\delta) \le H\(\frac 12 - \sqrt{\delta(1-\delta)}\)$, where $H(x)$ is the binary entropy function.

\myblt If $C$ is a linear code, then it is not hard to see that $\dim\(\Big<\big\{\Lambda_{r} \delta_x\big\}_{x \in C}\Big>\) = \frac{\Big |\bigcup_{z \in C^{\perp}} \big(z+B(r)\big) \Big |}{|C^{\perp}|}$, where $B(r)$ is the Hamming ball of radius $r$ around zero. Hence, for linear codes the claim of the theorem reduces to saying that the union of Hamming balls of radius $r(n)$ centered at the points of $C^{\perp}$ covers at least $\frac{1}{2d}$-fraction of the space. In this sense, the claim of the theorem is a proper generalization of the covering argument for linear codes given in \cite{ns1}.

\myblt The span $V_r$ of the Walsh-Fourier characters of weight at most $r$ is the space spanned by the eigenfunctions of the Laplacian operator on $\H$ corresponding to its smallest eigenvalues $0, 2, 4,... 2r$. In some metric spaces $X$ the spaces $V_r$ spanned by the eigenfunctions of the Laplacian corresponding to its lowest eigenvalues are the spaces of the ($r$-)'simple' functions on $X$. For instance, if $X$ is the Hamming cube $\H$, or the Euclidean sphere ${\mathbb S}^{n-1}$, then $V_r$ is the space of the real multivariate polynomials of degree at most $r$ on $\R^n$ restricted to either $\H$ or ${\mathbb S}^{n-1}$. One can ask for the value of $r$ for which the space $V_r$ becomes 'complex' enough to describe a given distance $d$ in the ambient space, in the sense that the projection of the characteristic function of any metric ball of
radius $d$ in the space on $V_r$ retains a significant fraction of its $\ell_2$ norm. For the Hamming cube $\H$ and $d = \delta n$, the appropriate value of $r$ is $\(\frac 12 - \sqrt{\delta(1-\delta)}\) \cdot n + o(n)$. (This is closely related to the fact the Krawchouk polynomial $K_d$ essentially attains its $\ell_2$ norm very close to its first root, but not much before that, see e.g., (11) and Proposition 2.15 in \cite{ks2}.) In other words, roughly speaking, the cardinality of a binary code of minimal distance $d$ is upperbounded by the dimension of the space $V_r$ spanned by the 'simple' eigenfunctions of the Laplacian if $r$ is large enough for the space $V_r$ to describe distances $d$ in $\H$. Let us remark that this phenomenon can also be shown to hold if the ambient space is the Hamming sphere (in effect recovering, via the Bassalygo-Elias inequality, the {\it second linear programming bound} for binary codes), and we believe that it should be possible to show this, by similar methods, for other symmetric spaces, such as distance regular graphs or the Eiclidean sphere. This seems to be rather intriguing, and we wonder whether this could be a special case of a more general principle.

\myblt The proof of Theorem~\ref{thm:JPL1} relies on the existence of a nonnegative function  $f$ on $\H$ with a small support (in fact it suffices to require that $\frac{\E f^2}{\E^2 f}$ is large), such that the adjacency matrix $A$ of the Hamming cube acts on $f$ by multiplying it pointwise by a large factor. It is not hard to see that any such function can be used to construct a feasible solution to the dual linear program of \cite{dels}.\footnote{Let $f \ge 0$ with $\frac{\E f^2}{\E^2 f} \ge \frac{2^n}{s}$, so that $Af \ge \l f$, for some $\l \ge 1$. Let $G = (A f) \ast f - (\l - 1) (f \ast f)$. It is easy to see that $\widehat{G}$ is a feasible solution to the dual program  of \cite{dels} for codes with minimal distance $d = \frac{n- \l + 1}{2}$, and the bound we get from the linear program is $A(n,d) \le s$.} In particular, if $f$ is the maximal eigenfunction of a Hamming ball, we (essentially) recover a solution to this program constructed in \cite{mrrw}.
In this sense, this line of research is subsumed by that of \cite{dels} and \cite{mrrw}. In addition, it can be shown \cite{sam-log-sob} that the best bound one can obtain following this approach is the first linear programming bound. With that, we believe that this approach leads to simpler proofs of this bound which provide additional information (we do not know how to derive the claim of the theorem from the linear program of \cite{dels}) and furthermore suggest new possible ways to proceed in order to improve this bound. In fact, we present two conjectures, suggested by the new proof, one for linear and one for general binary codes which, if true, would lead to an improvement of the first linear programming bound. We start with discussing the conjecture for linear codes.

\subsection{Linear codes}
\label{subsec:cnj-linear}

\noi Before stating the conjecture, let us mention two related conjectures that influenced it. Both conjectures posit, in different ways, that the behavior of a linear code near its minimal weight is highly constrained.

\noi The first conjecture, due to Kalai and Linial \cite{kal:lin}, states that if $C \subseteq \H$ is a linear code with minimal distance $d$, then the number of codewords of weight $d$ in $C$ is at most subexponential in $n$. In fact, they implicitly conjecture more, namely that there is also a strong upper bound on the number of codewords of weight close to $d$. They observe that if this is true, then the first linear programming bound for linear codes could be improved. This was elucidated in the subsequent work of \cite{abl}, where it was shown (among other things) that a linear code attaining the first linear programming bound must have (up to a negligible error) as many codewords of some weight close to $d$ as a random code of the same cardinality.

\noi A weak version of this conjecture, namely that the number of vectors of weight close to $d$ is exponentially smaller than the cardinality of $C$ (assuming $C$ is exponentially large, which is the interesting case here) was proved in \cite{ls}. However, the full conjecture was shown to be false in \cite{abv}, where a code of minimal distance $d$ with exponentially many codewords of weight $d$ was constructed.

\noi The second conjecture is due to H{\aa}stad \cite{Hastad-personal}. It states that for any absolute constants $0 < \alpha < 1$ and $k \ge 1$ there exists an absolute constant $K = K(\alpha,k)$, such that the following is true. Let $C$ be a {\it nice} (pseudorandom in some sense) linear code of length $n$ with minimal distance $d \le n^{\alpha}$. Then for any non-zero function $f$ on $\H$ whose Fourier transform is supported on vectors of weight at most $kd$ in $C$ holds
\beqn
\label{hastad}
\frac{\|f\|_4}{\|f\|_2} ~\le~ K,
\eeqn

\noi Some comments:

\myblt As pointed out in \cite{Hastad-personal} some precondition on $C$ is necessary. To see this, let $d =  \lfloor n^{\alpha} \rfloor$, and let $k = 3$. Let $m = 3d$, and let $C' \subseteq \{0,1\}^m$ be a linear code of minimal distance $d$ and dimension linear in $m$. Add $n-m$ zero coordinates to each vector in $C'$, obtaining a code $C \subseteq \H$. Take $f = \sum_{x \in C, |x| \le kd} W_x$ (here $\{W_x\}_x$ are the Walsh-Fourier characters of $\H$). Then it is easy to see that $f$ is essentially proportional to the characteristic function of the dual code $C^{\perp}$, and in particular $\frac{\|f\|_4}{\|f\|_2}$ is exponential in $m$.

\myblt The inequality (\ref{hastad}) is a {\it Khintchine-type inequality}. Recall that Khintchine-type inequalities establish an upper bound on the ratio of two $\ell_p$ norms for functions coming from a certain linear space, typically a space of multivariate polynomials of a specified degree over a given product space. In particular, the prototypical Khintchine inequality \cite{khintchine} states that the ratio of $\ell_2$ and $\ell_1$ norms of linear polynomials over the boolean cube $\H$ is bounded by an absolute constant. See \cite{it} for a recent discussion and references. 

\noi Our conjecture is in a sense a combination of the two conjectures above in that it considers the set of vectors in a linear code whose weight is close to the minimal distance of the code, but it replaces the 'hard' cardinality constraint of \cite{kal:lin} by a 'softer' analytical constraint ot \cite {Hastad-personal}.

\cnj
\label{cnj:main-linear}
Let $0 < \delta < \frac12$. There exists a positive constant $c = c(\delta)$, where one can take $c = \frac{\delta(1-2\delta) \cdot \log_2\(\frac{\frac12 + \sqrt{\delta(1-\delta)}}{\frac12 - \sqrt{\delta(1-\delta)}}\)}{16\sqrt{\delta(1-\delta)}}$, and a positive constant $\e_0 = \e_0(\delta)$ such that for any $0 \le \e \le \e_0$ the following holds. Let $C \subseteq \H$ be a linear code with minimal distance $d = \lfloor \delta n \rfloor$, let $d \le i \le (1+\e) d$, and let $A$ be the set of vectors of weight $i$ in $C$. Assume $A \not = \emptyset$. Let $f$ be a non-zero function on $\H$ whose Fourier transform is supported on $A$. Then
\beqn
\label{linear}
\frac{\|f\|_4}{\|f\|_2} ~\le~ 2^{c\e n + o(n)}.
\eeqn
\ecnj

\noi Some comments:

\myblt If $f$ is a non-zero function whose Fourier transform is supported on the vectors of minimal weight in a linear code, we conjecture that the ratio $\frac{\|f\|_4}{\|f\|_2}$ is at most subexponential in $n$. Since the ratio of the fourth and the second norm of a function is upper-bounded by the fourth root of the cardinality of its Fourier support (see e.g., Proposition 1.1 in \cite{ks1}) this conjecture is weaker than the corresponding conjecture of \cite{kal:lin}.

\myblt For weights close to minimal, the upper bounds on the ratio of the fourth and the second norm required in (\ref{linear}) are in general smaller than the explicit bounds on cardinality required in \cite{abl}. On the other hand, the quantity we want to bound in (\ref{linear}) is smaller, and in general could be much smaller (cf. Proposition~\ref{pro:LLC-random linear} in which for $i = (1+\e)d$ with $\e > 0$, the set of vectors of weight $i$ in the code is exponentially large, but the ratio of the moments is bounded by a constant). In this sense, the two conjectures are incomparable.

\myblt Conjecture~\ref{cnj:main-linear} and the conjecture of \cite{Hastad-personal} are also incomparable. Apart from the fact that the two conjectures speak about codes in different regimes (the conjecture of \cite{Hastad-personal} considers codes with sublinear distance, while we are interested in codes with linear distance), the conclusion of Conjecture~\ref{cnj:main-linear} has to hold for all linear codes, and not only for the 'nice' ones. On the other hand, the conclusion itself is much weaker, while the conditions under which it is supposed to hold are stronger.

\noi We claim that if Conjecture~\ref{cnj:main-linear} holds, then the first linear programming bound for binary linear codes can
be improved. In fact, it suffices to prove the conjecture only for {\it symmetric} functions $f$. Here we call function $f$ symmetric if for any point $x$ in its Fourier support, $\widehat{f}(x)$ depends only on $|x|$. Let $A_L(n,d)$ be the maximal size of a linear code of length $n$ and distance $d$. Let $R_L(\delta) = \limsup_{n \rarrow \infty} \frac 1n \log_2 A_L\(n,\lfloor \delta n \rfloor\)$.

\pro
\label{pro:linear-good}
Assume that  Conjecture~\ref{cnj:main-linear} holds for symmetric functions $f$. Then for all $0 < \delta < \frac12$ holds
\[
R_L(\delta) ~\le~  H\(\frac 12 - \sqrt{\delta(1-\delta)}\) - \theta_L(\delta),
\]
where $\theta_L(\delta) > 0$ for all $0 < \delta < \frac12$.
\epro

\noi Some comments: 

\myblt In the notation of Conjecture~\ref{cnj:main-linear}, if $f$ is symmetric, then $\widehat{f}$ is constant on $A$ and we may assume w.l.o.g. that $f = \sum_{a \in A} W_a$. With that, this special case essentially captures the complexity of the conjecture in full generality. In fact, it is known (see e.g., Proposition~1.1 in \cite{ks1}) that for any subset $A \subseteq \H$ the maximum of the ratio $\frac{\|f\|_4}{\|f\|_2}$ over non-zero functions $f$ whose Fourier transform is supported on $A$ is attained, up to a polylogarithmic in $|A|$ factor, on the characteristic function of some subset $B \subseteq A$. Hence, it would suffice to prove the conjecture for the linear code $C'$ spanned by the vectors in $B$ and for the appropriate symmetric function.

\myblt The fourth and the second norm in (\ref{linear}) may be replaced by any two norms $q > p$, changing the value of the constant $c(\delta)$ accordingly.

\noi We show that Conjecture~\ref{cnj:main-linear} holds (in a strong sense) for random linear codes. A random linear code $C$ of length $n$ and (prescribed) dimension $k$ is chosen as follows (see e.g., \cite{bf} for this and for properties of random linear codes): choose $k$ vectors $v_1,...,v_k$ independently at random from $\H$ and take $C$ to be the linear span of these vectors. It is convenient to define parameters in the following claim in terms of the dimension rather than the minimal distance of a code. This is justified by the following fact. Let $0 < R < 1$. A random linear code of dimension $k = \lfloor Rn \rfloor$ has minimal distance $H^{-1}(1-R) \cdot n \pm o(n)$ with probability tending to $1$ with $n$. So in case of random linear codes we may speak about the dimension of the code and its minimal distance interchangeably.

\pro
\label{pro:LLC-random linear}

\noi Let $0 < R < 1$. There exists a positive constant $K = K(R)$ and a positive constant $\e_0 = \e_0(R)$ such that the following holds with probability tending to $1$ with $n$ for a random linear code $C \subseteq \H$ of dimension $\lfloor Rn \rfloor$.  Let $d$ be the minimal distance of $C$, let $d \le i \le \(1+\e_0\) d$, and let $A$ be the set of vectors of weight $i$ in $C$. Assume $A \not = \emptyset$. Let $f$ be a function whose Fourier transform is supported on $A$. Then
\[
\frac{\|f\|_4}{\|f\|_2} ~\le~ K.
\]
\epro

\subsection{General codes}
\label{subsec:cnj-general}

\ignore{\noi We start with an introductory paragraph, whose goal is to explain the additional constraints on the code introduced in the following conjecture. The goal of this paragraph is to explain the additional constraints on the code introduced in the following conjecture.}

\noi We start with some definitions and a preliminary discussion.
\dfn
\label{dfn:i-graph}
For a binary code $C$ and $1 \le i \le n$ let $G(C,i)$, the distance-$i$ graph of $C$, be the graph with $|C|$ vertices indexed by the elements of $C$, with two vertices connected by an edge iff the corresponding elements of $C$ are at distance $i$ from each other.
\edfn

\noi If $C$ is a linear code, then $G(C,i)$ is a regular graph, for any $1 \le i \le n$. For a general code $C$ the graphs $G(C,i)$ could be highly irregular, which will lead to difficulties (see below) in formulating a conjecture for general codes analogous to Conjecture~\ref{cnj:main-linear}, which is not disproved by a simple counterexample. Consequently, we will need to introduce additional constraints in the conjecture below. Specifically, we will require the graphs $G(C,i)$ for $i$ close to the minimal distance of the code $C$ to behave somewhat similarly to regular graphs. In a regular graph $G$, the density of edges of any induced subgraph $G$ is bounded by that of the whole graph. We will call a graph $t$-balanced if it shares this property of regular graphs, up to a multiplicative factor of $t$.
\dfn
\label{dfn:balanced graph}
A non-empty graph $G = (V,E)$ will be called $t$-balanced, for some $t \ge 1$, if the density of edges in any of its induced subgraphs is at most $t$ times the density of edges in $G$. That is, denoting by $E(X,X)$ the set of edges from a subset $X$ of vertices to itself, for any $X \subseteq V$ holds $|\frac{|E(X,X)|}{|X|} \le t \cdot \frac{|E|}{|V|}$.
\edfn

\noi We can now state our conjecture for general graphs.

\cnj
\label{cnj:main-general}
Let $0 < \delta < \frac12$. There exists a positive constant $c = c(\delta)$, where one can take $c = \frac{\delta(1-2\delta) \cdot \log_2\(\frac{\frac12 + \sqrt{\delta(1-\delta)}}{\frac12 - \sqrt{\delta(1-\delta)}}\)}{16\sqrt{\delta(1-\delta)}}$, and a positive constant $\e_0 = \e_0(\delta)$, such that for any $0 \le \e \le \e_0$ the following holds. Let $C \subseteq \H$ be a code with minimal distance $d = \lfloor \delta n \rfloor$, let $d \le i \le (1+\e) d$, and let $G = G(C,i)$ be the distance-$i$ graph of $G$. Assume that $G$ is $t$-balanced, for $t = t(n) = 2^{o(n)}$. Let $\l \in \R^C$ be the vector of eigenvalues of $G$, viewed as a function on $C$, where we endow $C$ with the unform probability measure. Then
\beqn
\label{general}
\frac{\|\l\|_4}{\|\l\|_2} ~\le~ 2^{c\e n + o(n)}.
\eeqn
\ecnj

\noi Let us make several comments about this conjecture.

\myblt If $C$ is a linear code, then $G = G(C,i)$ is a regular graph, and hence it is $1$-balanced. Furthermore, if $A$ is the set of vectors of weight $i$ in $C$, then it is easy to see that the distribution of the eigenvalues of $G$ on $C$ is the same as the distribution of the function $f = \sum_{a \in A} W_a$ on $\H$ (see also the discussion in the proof of Proposition~\ref{pro:linear-good} below). Hence Conjecture~\ref{cnj:main-general} generalizes Conjecture~\ref{cnj:main-linear} for symmetric functions.

\myblt Some additional condition on $G = G(C,i)$ is necessary. Indeed, let $C$ be an exponentially large code with minimal distance $d$ which is linear in $n$. Let $k = \lfloor d/3 \rfloor$. Assume, w.l.o.g. that $k$ is even. Choose a point $x \in C$, and choose two points $y, z \in \H$ both at distance $k$ from $x$, such that the distance between $y$ and $z$ is also $k$ (clearly such points exist). Add $y$ and $z$ to $C$, obtaining a new code $C'$ with minimal distance $k$. It is easy to see that the graph $G(C,k)$ has only $3$ non-zero eigenvalues, and hence $\frac{\|\l\|_4}{\|\l\|_2} \ge \Omega\(|C|^{\frac14}\)$.

\myblt Recall that if $\l$ is the vector of eigenvalues of a graph $G$, then $\sum_i \l_i^k$ equals to the number of closed walks of length $k$ in $G$ (see e.g, Lemma 2.5 in \cite{biggs}). Hence Conjecture~\ref{cnj:main-general} can be informally restated as follows: Let $C$ be a binary code of length $n$ and distance $d$. Let $i$ be close to $d$ and assume that the graph $G(C,i)$ is $2^{o(n)}$-balanced (which always holds if $C$ is linear). Then the number of 'rhombic' $4$-tuples $\(x,y,z,w\) \in C^4$ with $|y-z| = |z-y| = |w-z| = |x-w| = i$ is not much larger then $|C| B^2_i$, where $B_i = \frac{1}{|C|} |\{(x,y) \in C \times C,~|x-y| = i\}$ is the $i^{th}$ component of the distance distribution vector of $C$.

\myblt Continuing from the preceding comment, we recall that the distance distribution vector $\(B_0,...,B_n\)$ of a code is a central object of study in the linear programming approach of \cite{dels}. The positive semidefinite approach of \cite{schrijver} (see also \cite{bgsv}) studies the geometry of a code in more detail, collecting the statistics of the possible ${k \choose 2}$-tuples of pairwise inner distances in all $k$-tuples of elements of a code, for some $k \ge 2$. In particular, statistics of inner distances of quadruples of codewords are studied in \cite{gms}. In this sense Conjecture~\ref{cnj:main-general} points out a possible connection between the two above-mentioned approaches whose eventual goal is to improve the linear programming bounds

\noi We claim that if Conjecture~\ref{cnj:main-general} holds, then the first linear programming bound for binary codes can
be improved.

\pro
\label{pro:general-good}
Assume that  Conjecture~\ref{cnj:main-general} holds. Then for all $0 < \delta < \frac12$ holds
\[
R(\delta) ~\le~  H\(\frac 12 - \sqrt{\delta(1-\delta)}\) - \theta(\delta),
\]
where $\theta(\delta) > 0$ for all $0 < \delta < \frac12$.
\epro

\noi We show that Conjecture~\ref{cnj:main-general} holds (in a strong sense) for random codes. See Section~\ref{subsec:backg} for more details on the (standard) model of random codes that we use. Let $0 < R < 1$. A random code of cardinality $2^{Rn}$ has minimal distance $H^{-1}(1-R) \cdot n \pm o(n)$ with probability tending to $1$ with $n$. So in case of random codes we may speak about the cardinality of the code and its minimal distance interchangeably.

\pro
\label{pro:LLC-random general}

\noi Let $0 < R < 1$. There exists a positive constant $K = K(R)$ and a positive constant $\e_0 = \e_0(R)$ such that the following holds with probability tending to $1$ with $n$ for a random code $C \subseteq \H$ of cardinality $2^{Rn}$.  Let $d$ be the minimal distance of $C$, let $d \le i \le \(1+\e_0\) d$, and let $G = G(C,i)$ be the distance-$i$ graph of $G$. Let $\l \in \R^C$ be the vector of eigenvalues of $G$, viewed as a function on $C$, where we endow $C$ with the unform probability measure. Then
\[
\frac{\|\l\|_4}{\|\l\|_2} ~\le~ K.
\]
\epro

\noi {\bf Organization of this paper}: The remainder of this paper is organized as follows. We describe the relevant notions and
provide some additional background in the next subsection. Theorem~\ref{thm:JPL1} is proved in Section~\ref{sec:JPL1}. Propositions~\ref{pro:linear-good}~and~\ref{pro:general-good} are proved in Section~\ref{sec:cnjs-good}. Propositions~\ref{pro:LLC-random linear}~and~\ref{pro:LLC-random general} are proved in Section~\ref{sec:LLC-random}.

\subsection{Background, definitions, and notation}
\label{subsec:backg}

\noi We view $\H$ as a metric space, with the Hamming distance between $x, y \in \H$ given by $|x - y| = |\{i: x_i \not = y_i\}|$. The Hamming weight of $x \in \H$ is $|x| = |\{i: x_i = 1\}|$. For $x, y \in \H$, we write $x+y$ for the modulo $2$ sum of $x$ and $y$. Note that the weight $|x+y|$ of $x+y$ equals to the distance $|x-y|$ between $x$ and $y$ (we will use this simple observation several times below).
The {\it Hamming sphere} of radius $r$ centered at $x$ is the set $S(x,r) = \left\{y \in \H:~|x-y| = r\right\}$. The {\it Hamming ball} of radius $r$ centered at $x$ is the set $B(x,r) = \left\{y \in \H:~|x-y| \le r\right\}$. Clearly, for any $x \in \H$ and $0 \le r \le n$ holds $|S(x,r)| = {n \choose r}$ and $|B(x,r) = \sum_{k=0}^r {n \choose k}$.

\noi Let $H(t) = t \log_2\(\frac 1t\) + (1-t) \log_2\(\frac{1}{1-t}\)$ be the binary entropy function. We will make use of the following estimate (see e.g., Theorem~1.4.5. in \cite{van lint}): For $x \in \H$ and $0 < r \le \frac n2$ holds
\beqn
\label{binom}
|B(x,r)| ~\le~ 2^{H\(\frac{r}{n}\) \cdot n}.
\eeqn

\noi The asymptotic notation will always refer to the behavior of a function of an integer argument $n$ when $n$ tends to infinity (unless specifically stated otherwise). The $O$, $\Omega$ and $\Theta$ asymptotic notation always hides absolute constants. 

\noi We write $a \in b \pm \e$ as a shorthand for $b - \e \le a \le b + \e$.

\subsubsection{Fourier analysis, Krawchouk polynomials, and spectral projections}
\label{subsubsec:Fourier}

\noi We recall some basic notions in Fourier analysis on the boolean cube (see \cite{O'Donnel}). For $\alpha \in \H$, define the Walsh-Fourier character $W_{\alpha}$ on $\H$ by setting $W_{\alpha}(y) = (-1)^{\sum \alpha_i y_i}$, for all $y \in \H$. The {\it weight} of the character $W_{\alpha}$ is the Hamming weight $|\alpha|$ of $\alpha$.  The characters $\{W_{\alpha}\}_{\alpha \in \H}$ form an orthonormal basis in the space of real-valued functions on $\H$, under the inner product $\<f, g\> = \frac{1}{2^n} \sum_{x \in \H} f(x) g(x)$. The expansion $f = \sum_{\alpha \in \H} \widehat{f}(\alpha) W_{\alpha}$ defines the Fourier transform $\widehat{f}$ of $f$. We also have the Parseval identity, $\<f,g\> = \sum_{\alpha \in \H} \widehat{f}(\alpha) \widehat{g}(\alpha)$. We will write the RHS of this identity (the inner product in the "Fourier domain") as $\<\widehat{f},\widehat{g}\>_{\cal F}$.
The {\it convolution} of $f$ and $g$ is defined by $(f \ast g)(x) = \frac{1}{2^n} \sum_{y \in \H} f(y) g(x+y)$. The convolution transforms to dot product:  $\widehat{f\ast g} = \widehat{f} \cdot \widehat{g}$. The convolution operator is commutative and associative. We will use one additional simple fact. Let $A$ be the adjacency matrix of $\H$. Then $\widehat{Af}(\alpha) = (n - 2\alpha) \widehat{f}(\alpha)$.

\noi {\it Krawchouk polynomials}. For $0 \le s \le n$, let $F_s$ be the sum of all Walsh-Fourier characters of weight $s$, that is $F_s = \sum_{|\alpha| = s} W_{\alpha}$. Note that $F_s$ is the Fourier transform of $2^n \cdot L_s$, where $L_s$ is the characteristic function of the Hamming sphere of radius $s$ around $0$. It is easy to see that $F_s(x)$ depends only on the Hamming weight $|x|$ of $x$, and it can be viewed as a univariate function on the integer points $0,...,n$, given by the restriction to $\{0,...,n\}$ of the univariate polynomial $K_s = \sum_{k=0}^s (-1)^k {x \choose k} {{n-x} \choose {s-k}}$ of degree $s$. That is, $F_s(x) = K_s(|x|)$. The polynomial $K_s$ is the $s^{th}$ {\it Krawchouk polynomial}. Abusing notation, we will also call $F_s$ the $s^{th}$ Krawchouk polynomial, and write $K_s$ for $F_s$ when the context is clear.

\noi {\it Spectral projections}. For $0 \le r \le n$ we define $\Lambda_r$ to be the orthogonal projection to the subspace spanned by Walsh-Fourier characters of weight at most $r$. That is, for a function $f$ on $\H$, and $0 \le r \le n$, we have $\Lambda_r f = \sum_{|\alpha| \le r} \widehat{f}(\alpha) W_{\alpha}$.

\subsubsection{Bounds on the asymptotic rate function}
\label{subsubsec:rate bounds}

\noi The best known lower bound on $R(\delta)$ is the {\it Gilbert-Varhsamov bound} $R(\delta) \ge 1 - H(\delta)$ (see e.g., \cite{macw-sl}). The existence of codes asymptotically attaining this bound is demonstrated e.g., by random codes (see Section~\ref{subsubsec:random}). The best known upper bounds on $R(\delta)$ are the {\it linear programming bounds} \cite{mrrw}, obtained via the linear programming approach of \cite{dels}. To be more specific, \cite{dels} suggested a systematic approach to obtaining a linear programming relaxation of the combinatorial problem of bounding the cardinality of an error-correcting code with a given minimal distance (equivalently, of a metric ball packing with a given radius of a ball) in a metric space with a large group of isometries and, more generally, in an association scheme.
In \cite{mrrw} tools from the theory of orthogonal polynomials were used to construct good feasible solutions to the linear programs constructed in \cite{dels} for the Hamming cube and the Hamming sphere. This led to two families of bounds. The first linear programing bound is obtained by solving the linear program for the cube. Solving the linear program for the sphere gives bounds for codes in the sphere (also known as constant weight codes). Bounds on codes in the sphere lead to bounds for codes in the cube, via the Bassalygo-Elias inequality, reflecting fact that the sphere (of an appropriate dimension) is a subset of the cube. Optimizing over the radius of the embedded sphere leads to the second linear programing bound. The two bounds coincide for relative distance $\delta \ge 0.273...$. In the remaining range the second bound is better.

\subsubsection{Random codes}
\label{subsubsec:random}

\noi There are standard models of random binary codes (see e.g., \cite{bf}). A {\it random linear code} $C$ of length $n$ and (prescribed) dimension $k$ is chosen as follows: choose $k$ vectors $v_1,...,v_k$ independently at random from $\H$ and take $C$ to be the linear span of these vectors. If $k = \lfloor Rn \rfloor$ for some $0 < R < 1$, then the following two events hold with probability tending to $1$ with $n$. The code $C$ has dimension $k$ and minimal distance $d \in H^{-1}(1-R) \cdot n \pm o(n)$. {\it Shorthand}: Here and below we will write 'with high probability' (w.h.p.) as a shorthand for 'with probability tending to $1$ with $n$'.

\noi We pass to nonlinear codes. Let $0 < R < 1$. A {\it random code} $C$ of (prescribed) cardinality $N = \lfloor 2^{Rn} \rfloor$ is chosen in two steps. First, we choose $N$ points $x_1,...,x_N$ independently at random from $\H$. Next, we erase from this list pairs of points whose distance from each other lies below a certain threshold. There are several essentially equivalent ways to choose this threshold. The model we use is as follows: Fix a sufficiently small (see the discussion before the proof of Proposition~\ref{pro:LLC-random general}) constant $\tau = \tau(R)$, and define $d_0$ to be the maximal integer between $1$ and $N$ so that $\frac{N}{2^n} \sum_{\ell = 0}^{d_0-1}  \cdot {n \choose {\ell}} \le \tau$. For all $1 \le i <  j \le N$ such that $|x_i - x_j| \le d_0-1$, erase the points $x_i$ and $x_j$ from the list $x_1,...,x_N$. Take $C$ to be the remaining collection of points. The following two events hold with high probability: $C \ge \Omega(N)$ and the minimal distance of $C$ is in $H^{-1}(1-R) \cdot n \pm o(n)$.

\section{Proof of Theorem~\ref{thm:JPL1}}
\label{sec:JPL1}

\noi We consider a code $C$ of length $n$ and distance $d = \lfloor \delta n \rfloor$. In the first part of the argument we find a nonnegative function $\phi$ on $\H$ with small support such that the adjacency matrix $A$ of $\H$ acts on $\phi$ by multiplying it pointwise by the factor of at least $n-2d+1$. We chose this function to be the maximal eigenfunction of the Hamming ball of an appropriate radius around zero. For $0 \le r \le n$, let $B_r$ be the Hamming ball of radius $r$ around $0$ in $\H$. Let $A_r$ be the adjacency matrix of the subgraph of $\H$ induced by the vertices of $B_r$ and let $\l_r$ be the maximal eigenvalue of $A_r$. Clearly, $\l_r$ is an increasing function of $r$, with $\l_0 = 1$ and $\l_n = n$. Let $r_0$ be the smallest value of $r$ for which the maximal eigenvalue of $\Lambda_r$ is at least $n-2d+1$. It was shown in \cite{ft} (see also Lemma~3.3 in \cite{ns} for a direct argument) that $\l_r \ge 2\sqrt{r(n-r)} - o(n)$. Hence $r_0 \le \(\frac12 - \sqrt{\delta(1-\delta)}\) \cdot n + o(n)$. Abusing notation, we write $r$ for $r_0$ from now on.

\noi Let $\phi = \phi_r$ be the maximal eigenfunction of $A_r$ with $\|\phi\|_2 = 1$. Since $B_r$ is a connected graph which is invariant under permutations of the coordinates, the function $\phi$ is uniquely defined. It is positive on $B_r$ and symmetric ($\phi(S) = \phi(|S|)$, for $0 \le |S| \le r$). We extend $\phi$ to the whole space $\H$ by setting $\phi = 0$ outside $B_r$ and, abusing notation, write $\phi$ for this extension as well. We record the relevant properties of $\phi$:

\begin{enumerate}

\item $\phi$ is supported on $B_r$.

\item $\phi$ is nonnegative and symmetric.

\item $Af \ge \l_r \cdot \phi \ge (n-2d+1) \cdot \phi$, with all inequalities holding pointwise on $\H$.

\end{enumerate}

\noi We use these properties of $\phi$ to show that for the orthogonal projection $\Lambda_r$ on the span of the Walsh-Fourier characters of weight at most $r$ holds $\dim\(\Big<\big\{\Lambda_{r(n)} \delta_x\big\}_{x \in C}\Big>\) \ge \frac{1}{2d} \cdot |C|$, proving the claim of the theorem.

\noi Let the matrix $M = M_r$ be defined as follows. The rows of $M$ are indexed by the elements of $C$ and the columns by the subsets of $[n]$ of cardinalities $0...r$, arranged in increasing order of cardinalities. For $y \in C$ and $S \subseteq [n]$, let $M(y,S) = W_S(y) = (-1)^{\<y,S\>}$ (viewing $S$ as an element of $\H$). Observe that the row of $M$ indexed by $x \in C$ contains the non-vanishing part of the Fourier expansion of $2^n \cdot \Lambda_r\(\delta_x\)$, and hence the rank of $M$ equals to $\dim\(\Big<\big\{\Lambda_{r(n)} \delta_x\big\}_{x \in C}\Big>\)$. Let $D$ be the $|B_r| \times |B_r|$ diagonal matrix indexed by the subsets of $[n]$ of cardinality at most $r$, with $D(S,S) = \phi(S)$ for all $|S| \le r$, and let ${\cal M} = M D M^t$. Then ${\cal M}$ is a $|C| \times |C|$ matrix whose rank is the same as the rank of $M$. Hence it suffices to show that the rank of ${\cal M}$ is at least $\frac{1}{2d} \cdot |C|$. For the remainder of the proof we write $N$ for $|C|$, for typographic convenience.

\noi Let $\l_1...\l_N$ be the eigenvalues of ${\cal M}$. We will show that $2d \cdot \(\frac1N \sum_{i=1}^N \l_i\)^2 \ge \frac1N \sum_{i=1}^N \l_i^2$. This will imply, by the Cauchy-Schwarz inequality, that the number of non-zero eigenvalues is at least $N / 2d$, proving the claim.
Writing ${\cal M} = \(m_{y,z}\)_{y,z \in C}$, we can write this inequality in terms of the entries of ${\cal M}$: $2d  \cdot \frac{1}{N^2} \(\sum_{y \in C} m_{y,y}\)^2 \ge \frac{1}{N} \sum_{y,z \in C} m^2_{y,z}$.

\noi Let $f = 2^n \cdot \widehat{\phi}$. By the definition of ${\cal M}$, for $y, z \in C$ holds
$m_{y,z} = \sum_{S:|S| \le r} \phi(S) (-1)^{\<y+z,S\>} = f(y+z)$. So, we need to show that
\beqn
\label{M-rank}
\Big(2d\Big) \cdot  N f^2(0) ~\ge~ \sum_{y,z \in C} f^2\(y + z\).
\eeqn

\noi To do this we estimate $\<\Big(A \phi\Big) \ast \phi,~ \widehat{1_C}^2\>_{\cal F}$ in two ways. On one hand,
\[
\<\Big(A \phi\Big) \ast \phi,~ \widehat{1_C}^2\>_{\cal F} ~\ge~ (n-2d+1) \cdot \<\phi \ast \phi, ~\widehat{1_C}^2\>_{\cal F} ~=~ (n-2d+1) \cdot \<f^2, ~1_C \ast 1_C\> ~ =
\]
\[
\frac{n-2d+1}{2^{2n}} \cdot \sum_{y, z \in C} f^2(y+z).
\]
We used the first and the second properties of $\phi$ in the first step, and Parseval's identity in the second step. Note that $\phi = \widehat{f}$. On the other hand,
\[
\<\Big(A \phi\Big) \ast \phi,~ \widehat{1_C}^2\>_{\cal F} ~=~ \<(n-2|x|)  \cdot f(x)^2,~1_C \ast 1_C \> ~=~ \frac{1}{2^{2n}} \sum_{y,z \in C} \Big(n - 2|y+z|\Big) ~f^2(y+z) ~\le
\]
\[
\frac{n}{2^{2n}} \cdot N  f^2(0) + \frac{n - 2d}{2^{2n}} \cdot \sum_{y \not = z \in C} f^2\(y + z\) ~=~
\frac{2d}{2^{2n}} \cdot N  f^2(0) ~+~ \frac{n - 2d}{2^{2n}} \cdot \sum_{y, z \in C} f^2\(y + z\).
\]

\noi We used Parseval's identity in the first step, and the fact that $C$ has distance $d$ in the third step. Combining the two estimates and simplifying gives (\ref{M-rank}). \eprf

\section{Proof of Propositions~\ref{pro:linear-good} and \ref{pro:general-good}}
\label{sec:cnjs-good}

\noi The proofs will follow in outline the proof of Theorem~\ref{thm:JPL1}. Somewhat imprecisely speaking, if either of Conjectures~\ref{cnj:main-linear}~or~\ref{cnj:main-general} holds, we would be able to replace the Hamming ball of radius $r(n) = \(\frac12 - \sqrt{\delta(1-\delta)}\) \cdot n + o(n)$ in the argument with a Hamming ball of a significantly smaller radius. We use the same notation as in the proof of Theorem~\ref{thm:JPL1}.

\noi We start with a technical lemma, which provides a useful description of the symmetric function $f_r = 2^n \widehat{\phi_r}$. See Section~\ref{subsubsec:Fourier} for the relevant notions in Fourier analysis on $\H$.

\lem
\label{lem:eigen-Kr}
Let $0 \le r < n$. Then, for any $x \in \H$ with $|x| \not = \frac{n-\l_r}{2}$ holds
\[
f_r(x) ~=~ c \cdot \frac{K_{r+1}(x)}{n - \l_r - 2|x|},
\]
where $K_{r+1}$ is the appropriate Krawchouk polynomial, and $c$ is a positive constant.
\elem

\prf
We view $\phi_r$ as a function on $\H$. Since it is symmetric and supported on $B_r$, we can write $\phi_r = \sum_{i=0}^r a_i L_i$, where $L_i$ is the characteristic function of the Hamming sphere of radius $i$ around zero, and the coefficients $a_i$ are positive. Note that $A \phi_r = \l_r \cdot \phi_r + (r+1) a_r L_{r+1}$. Multiplying both sides of this equality by $2^n$ and applying the Fourier transform we have, for $x \in \H$:
$(n-2|x|) \cdot f(x) ~=~ \l_r \cdot f(x) + (r+1) a_r K_{r+1}(x)$, which implies the claim of the lemma.
\eprf

\cor
\label{cor:eigen-root}
\begin{itemize}

\item $\frac{n-\l_r}{2}$ is a root of $K_{r+1}$, with $K_{r+1}$ viewed as the appropriate univariate real polynomial of degree $r+1$.

\item $\l_r \in 2\sqrt{r(n-r)} \pm o(n)$.
\end{itemize}
\ecor
\prf

\noi In the notation of the proof of Lemma~\ref{lem:eigen-Kr}, we have $f_r = \sum_{i=0}^r a_i K_i$. Viewed as a univariate polynomial, this is a polynomial of degree $r$, and we have the identity $(n - \l_r - 2k) f_r(k) = K_{r+1}(k)$, for all integer $k$ between $0$ and $n$. This means that $(n - \l_r - 2x) \cdot f_r(x) = K_{r+1}(x)$ for all real $x$, implying that $\frac{n-\l_r}{2}$ is a root of $K_{r+1}$.

\noi In particular, $\l_r \le n - 2x_{r+1}$, where $x_{r+1}$ is the minimal root of $K_{r+1}$.  Using the known estimates on $x_{r+1}$ (see e.g., \cite{lev-chapter}) gives $\l_r \le 2\sqrt{r(n-r)} - o(n)$. The second claim of the corollary follows from this and from the estimate $\l_r \ge 2\sqrt{r(n-r)} + o(n)$ \cite{ft}.
\eprf

\subsection{Proof of Proposition~\ref{pro:linear-good}}

\noi Fix $0 < \delta < \frac12$ and assume that Conjecture~\ref{cnj:main-linear} holds for this value of $\delta$. For an integer $n$, let $d = \lfloor \delta n \rfloor$. Let $C$ be a linear code of length $n$ and distance $d$. Let $r = r(n)$ be the minimal radius of a Hamming ball centered at $0$ for which $\l_r \ge n - 2(1+\e)d$, where $\e \le \e_0$ and $\e_0 = \e_0(\delta)$ is the constant specified by Conjecture~\ref{cnj:main-linear}. We proceed as in the proof of Theorem~\ref{thm:JPL1}, using  the same notation, but replacing the value $r = r_0$ in that proof with the new value of $r$ we have chosen. Computing $\<\Big(A \phi\Big) \ast \phi,~ \widehat{1_C}^2\>_{\cal F}$ in two ways, we get,on one hand,
\[
\<\Big(A \phi\Big) \ast \phi,~ \widehat{1_C}^2\>_{\cal F} ~\ge~ \l_r \cdot \<\phi \ast \phi, ~\widehat{1_C}^2\>_{\cal F} ~=~ \l_r \cdot \<f^2, ~1_C \ast 1_C\> ~ =
\]
\[
\frac{\l_r}{2^{2n}} \cdot \sum_{y, z \in C} f^2(y+z).
\]

\noi On the other hand, assuming w.l.o.g. that $(1+\e) d$ is an integer (which we can with a negligible loss), we have
\[
\<\Big(A \phi\Big) \ast \phi,~ \widehat{1_C}^2\>_{\cal F} ~=~ \<(n-2|x|)  \cdot f(x)^2,~1_C \ast 1_C \> ~=~ \frac{1}{2^{2n}} \sum_{y,z \in C} \Big(n - 2|y+z|\Big) ~f^2(y+z) ~\le
\]
\[
\frac{n}{2^{2n}} \cdot N  f^2(0) + \frac{n - 2d}{2^{2n}} \cdot \sum_{y \not = z \in C, |y-z| \le (1+\e) d} f^2\(y + z\) + \frac{\l_r - 2}{2^{2n}} \cdot \sum_{y, z\in C, |y-z| > (1+\e) d} f^2\(y + z\).
\]

\noi Combining both estimates, rearranging, and writing for simplicity a larger expression than needed on the LHS of the following inequality, we get
\[
n \cdot N  f^2(0) + n \cdot \sum_{y \not = z \in C, |y-z| \le (1+\e) d} f^2\(y + z\) ~\ge~ \sum_{y, z\in C} f^2\(y + z\).
\]

\noi This means that one of the summands on the LHS is at least as large as $\frac12 \cdot \sum_{y, z\in C} f^2\(y + z\)$. We consider both of these possibilities.

\begin{enumerate}

\item $n \cdot N  f^2(0) \ge \frac12 \cdot \sum_{y, z\in C} f^2\(y + z\)$.

\noi In this case, we can proceed as in the proof of Theorem~\ref{thm:JPL1} and deduce that the rank of the matrix ${\cal M}$ is at least $\frac{1}{2n} \cdot |C|$. On the other hand, this rank is at most $\sum_{i=0}^r H(i) \le 2^{H\(\frac rn\) \cdot n}$ (where we have used (\ref{binom})). Hence $|C| \le 2n \cdot 2^{H\(\frac rn\) \cdot n}$. By the definition of $r$ and by the second claim of Corollary~\ref{cor:eigen-root} we get that $r \le \(\frac12 - \sqrt{\delta(1-\delta)}\) \cdot n  - an$, for some absolute constant $a = a(\delta)$. Hence we get an upper bound on $|C|$ which is exponentially smaller than the first linear programming bound, completing the proof of the proposition in this case.

\item $n \cdot \sum_{y \not = z \in C, |y-z| \le (1+\e) d} f^2\(y + z\) ~\ge~ \frac12 \cdot \sum_{y, z\in C} f^2\(y + z\)$.

\noi This means that for some $d \le i \le (1+\e) d$ holds $n^2 \cdot \sum_{y,z \in C, |y-z| = i} f^2\(y + z\) \ge \sum_{y, z\in C} f^2\(y + z\)$. Let $A$ be the $|C| \times |C|$ matrix indexed by the elements of $C$ with $A(y,z) = \begin{cases} f(y+z) & |y-z| = i\\ 0 & \mathrm{otherwise} \end{cases}$. Then the preceding inequality can be written as $n^2 \cdot Tr\(A{\cal M}\) \ge Tr\({\cal M}^2\)$.

\noi Let $F = \sum_{x \in |C|,\,|x|=i} f(x) W_x$. Since $C$ is a linear code, it is  well-known (and easy to see) that the eigenvectors of $A$ are the restrictions to $C$ of the Walsh-Fourier characters $\{W_u\}_{u \in \H}$, and the eigenvalue corresponding to $W_u$ is $F(u)$. Since the restrictions of $W_u$ and $W_{u'}$ coincide iff $u$ and $u'$ are in the same coset of $C^{\perp}$, the function $F$ is constant on the cosets of $C^{\perp}$ in $\H$, and the distribution of $F$ in $\H$ is the same as the distribution of the eigenvalues of $A$ in $C$, provided both $\H$ and $C$ are endowed with uniform probability measure. Let $\alpha = \(\alpha_1,...,\alpha_N\)$ be the vector of eigenvalues of $A$, viewed as a function on $C$. By the preceding discussion, we have that $\frac{\|\alpha\|_4}{\|\alpha\|_2} = \frac{\|F\|_4}{\|F\|_2}$.

\noi On the other hand, $F$ is a symmetric function whose Fourier transform is supported on the set of vectors of weight $i$ in $C$. The conditions of Conjecture~\ref{cnj:main-linear} are satisfied, and since we have assumed the conjecture to hold we have $\frac{\|\alpha\|_4}{\|\alpha\|_2} = \frac{\|F\|_4}{\|F\|_2} \le 2^{c \e n + o(n)}$, where $c = c(\delta)$ is the constant specified in the conjecture. 

\noi Let $\l = \(\l_1,...,\l_N\)$ be the vector of eigenvalues of ${\cal M}$. Assume that both $\alpha$ and $\l$ are arranged in decreasing order of values. Note that $Tr\({\cal M}^2\) = \|\l\|_2^2$. We also have $Tr\(A{\cal M}\) \le \<\alpha, \l\>$ by \cite{t}.\footnote{Since $C$ is a linear code, the matrices $A$ and ${\cal M}$ commute, so the result of \cite{t} is not required. With that we state the argument in higher generality, to apply to general codes as well.} Hence $n^2 \cdot Tr\(A{\cal M}\) \ge Tr\({\cal M}^2\)$ implies $n^2 \cdot \<\alpha, \l\> \ge \|\l\|_2^2$.

\noi Taking everything into account, we have
\[
\|\l\|_2^2 ~\le~ n^2 \cdot \<\alpha, \l\> ~\le~ n^2 \cdot \|\alpha\|_4 \|\l\|_{\frac43} ~\le~ 2^{c \e n + o(n)} \cdot \|\alpha\|_2 \|\l\|_{\frac43} ~\le~ 2^{c \e n + o(n)} \cdot \|\l\|_2 \|\l\|_{\frac43},
\]
where in the second step we use H\"older's inequality, and in the last step the elementary fact $\|\alpha\|_2 \le \|\l\|_2$. So we get $\|\l\|_2 \le 2^{c \e n + o(n)} \cdot \|\l\|_{\frac43}$. Let $S \subseteq C$ be the support of $\l$. Note that $|S| = \mathrm{rank}({\cal M})$. Applying H\"older's inequality once again (in the second step below), we have
\[
\|\l\|_{\frac43}^{\frac43} ~=~ \<\l^{\frac43}, 1_S\> ~\le~ \|\l^{\frac43}\|_{\frac 32} \|1_S\|_3 ~=~ \|\l\|_2^{\frac43} \(\frac{|S|}{|C|}\)^{\frac13},
\]
implying that $\mathrm{rank}({\cal M}) = |S| \ge |C| \cdot \(\frac{\|\l\|_{\frac43}}{\|\l\|_2}\)^4 \ge |C| \cdot 2^{-4c \e n + o(n)}$. On the other hand, we have $\mathrm{rank}({\cal M}) = \mathrm{rank}(M) \le \sum_{i=0}^r {n \choose i} \le 2^{H\(\frac rn\) \cdot n}$. So, we get
\[
\frac 1n \log_2 |C| \le H\(\frac rn\) +4c\e + o(1).
\]
It remains to analyze the expression on the RHS of this inequality. Let $\rho = \frac rn$, let $\rho_0 = \frac12 - \sqrt{\delta(1-\delta)}$, and let $g(x) = 2\sqrt{x(1-x)}$. Note that $g\(\rho_0\) = 1 - 2\delta$. Ignoring negligible factors, which we do from now on in this calculation, we also have $g(\rho) = 1 - 2\delta - 2\delta \e$. Assuming $\e$ is sufficiently small, and using first order approximations, we have $\rho \approx \rho_0 - \frac{2\delta \e}{ g'\(\rho_0\)}$, and hence $H\(\rho\) \approx H\(\rho_0\) - \frac{H'\(\rho_0\)}{ g'\(\rho_0\)} \cdot 2\delta \e$. It is easy to verify that $c = c(\delta) =  \frac18 \cdot \frac{H'\(\rho_0\)}{ g'\(\rho_0\)} \cdot 2\delta$, and hence
\[
H\(\frac rn\) +4c\e ~\approx~ H\(\rho_0\) - 4c\e ~=~ H\(\frac12 - \sqrt{\delta(1-\delta)}\) - 4c\e,
\]
completing the proof the proposition in this case.

\end{enumerate}

\eprf

\subsection{Proof of Proposition~\ref{pro:general-good}}

\noi Let $0 < \delta < \frac12$ and assume that Conjecture~\ref{cnj:main-general} holds for this value of $\delta$. We will assume that there is a sequence of codes $C_n$ of length $n$ and distance $d = \lfloor \delta n \rfloor$ attaining the first linear programming bound and reach a contradiction. Assume then that $n$ is large, and that $C$ is a code of length $n$, distance $d = \lfloor \delta n \rfloor$, such that $\frac 1n \log_2 |C| \ge H\(\frac12 - \sqrt{\delta(1-\delta)}\) - o(1)$. 

\noi We follow the same argument as in the proof of Proposition~\ref{pro:linear-good}. It is readily seen that everything works through if we show that if $d \le i \le (1+\e) d$ is such that $n^2 \cdot \sum_{y,z \in C, |y-z| = i} f^2\(y + z\) \ge \sum_{y, z\in C} f^2\(y + z\)$, then the vector $\alpha$ of eigenvalues of the distance-$i$ graph $G(C,i)$ satisfies $\frac{\|\alpha\|_4}{\|\alpha\|_2} \le 2^{c \e n + o(n)}$. This will follow from Conjecture~\ref{cnj:main-general} if we show that $G(C,i)$ is $2^{o(n)}$-balanced. This is what we proceed to show. We start with a technical lemma.

\lem
\label{lem:f-technical}
Let $K_s$ be a Krawchouk polynomial, for some $1 \le s \le \frac n2$. Let $a$ be a root of $K_s$, and let $g = \frac{K_s}{x-a}$. Then 
\begin{enumerate}

\item $\|g\|_2^2 = \frac{1}{2^n} \sum_{i=0}^n {n \choose i} g^2(i) \ge \frac{1}{n^2} \cdot {n \choose s}$.

\item For all $0 \le k \le n$ holds $g^2(k) \le O\(n^3\) \cdot \frac{2^n {n \choose s}}{{n \choose k}}$. 

\end{enumerate}
\elem

\prf
Recall (see \cite{lev-chapter} for this and for additional properties of Krawchouk polynomials used in this proof) that $\|K_s\|_2^2 = {n \choose s}$, and that all the roots of $K_s$ lie in the interval $(0,n)$. Hence $|k - a| < n$ for all $0 \le k \le n$, and first claim of the lemma follows.

\noi We pass to the second claim of the lemma. It is known that $K_s$ has $s$ simple roots. Let them be $x_1 < x_2 < ... < x_s$, and let $a = x_m$, for some $1 \le m \le s$. There are two cases to consider. Either $k$ lies inside the root region of $K_s$, that is $x_1 \le k \le x_s$, or not. We consider the first case. (The second case is similar and simpler.) Since $\|K_s\|_2^2 = {n \choose s}$, for all $0 \le i \le n$ holds $K_s^2(i) \le \frac{2^n {n \choose s}}{{n \choose i}}$. Hence if $|k - a| > \frac12$ the claim follows immediately. If $|k - a| < \frac 12$ there are again two cases to consider, $k > a$ and $k < a$. We consider the first case, the second is similar. Recall that the distance between any two consecutive roots of $K_s$ is at least $2$. Since $a = x_m \le k < x_m + \frac 12$, this means that the point $k+1$ lies between $x_m$ and $x_{m+1}$ and it is at distance at least $\frac 12$ from $x_{m+1}$. This implies that
\[
\Big | \frac{g(k)}{g(k+1)} \Big | ~=~ \prod_{\ell \not = m} \frac{|k-x_{\ell}|}{|k+1-x_{\ell}|} ~\le~ \frac{x_s - x_m}{x_{m+1}-k-1} ~\le~ 2n.
\]

\noi It follows that
\[
g^2(k) ~\le~ 4n^2 \cdot g^2(k+1) ~\le~ 4n^2 \cdot \frac{2^n {n \choose s}}{{n \choose {k+1}}} ~\le~ 4n^3 \cdot \frac{2^n {n \choose s}}{{n \choose k}}.
\]
\eprf

\noi We proceed with the argument. Let $\(B_0,...,B_n\)$ be the distance distribution vector of $C$, with $B_k = \frac{1}{|C|} |\{(x,y) \in C \times C,\,|x-y| = k\}|$. Note that 
\[
\cdot \sum_{y,z \in C, |y-z| = i} f^2\(y + z\) ~=~ |C| B_i f^2(i) ~\le~ O\(n^3\) \cdot |C|B_i \cdot \frac{2^n {n \choose {r+1}}}{{n \choose i}},
\]
where in the second step we have used Lemma~\ref{lem:eigen-Kr}, the first claim of Corollary~\ref{cor:eigen-root}, and the second claim of Lemma~\ref{lem:f-technical}. On the other hand, we have
\[
\sum_{y,z \in C} f^2(y+z) ~=~ 2^n \cdot \<1_C \ast 1_C, f^2\> ~=~ 2^{2n} \cdot \<\widehat{1_C}^2, \widehat{f} \ast \widehat{f}\>_{\cal F} ~\ge~ 
\]
\[
\(2^n \widehat{1_C}(0)\)^2 \cdot (f \ast f)(0) ~=~ |C|^2 \|f\|_2^2 ~\ge~ \frac{1}{n^2} |C|^2 {n \choose {r+1}},
\]
where we have used Parseval's identity in the second step, the fact that $\widehat{f} = \phi$ is nonnegative in the third step, and the first claim of Lemma~\ref{lem:f-technical} in the fifth step. This means that $n^2 \cdot \sum_{y,z \in C, |y-z| = i} f^2\(y + z\) \ge \sum_{y, z\in C} f^2\(y + z\)$ implies
\[
B_i ~\ge~ \Omega\(\frac{1}{n^5}\) \cdot \frac{|C| {n \choose i}}{2^n} ~\ge~ 2^{-o(n)} \cdot 2^{H\(\frac12 - \sqrt{\delta(1-\delta)}\) \cdot n} \cdot \frac{{n \choose i}}{2^n},
\]
where we have used the assumption that $C$ attains the first linear programming bound in the second step. Note that this means that the edge density of the graph $G(C,i)$ is at least $2^{-o(n)} \cdot 2^{H\(\frac12 - \sqrt{\delta(1-\delta)}\) \cdot n} \cdot \frac{{n \choose i}}{2^n}$.

\noi (Observe that the preceding discussion provides an additional proof of the fact (\cite{abl}) that a linear code attaining the first linear programming bound must have (up to a negligible error) as many codewords of some weight close to its minimal distance as a random code of the same cardinality, see Section~\ref{subsec:cnj-linear}. It is in fact possible that the above inequality for $B_i$ might have been derived directly from Corollary~1 in \cite{abl}, but we have not found a ready way to do so.)

\noi Corollary~1 in \cite{abl} also presents a complementary result: Let $C'$ be a code with distance $d = \lfloor \delta n \rfloor$, and with distance distribution $\(B'_0,...,B'_n\)$. Then for any $d \le k \le n/2$ holds $B'_k \le 2^{o(n)} \cdot 2^{H\(\frac12 - \sqrt{\delta(1-\delta)}\) \cdot n} \cdot \frac{{n \choose k}}{2^n}$. 

\noi Let now $C'$ be a subset of $C$, and consider the subgraph $G\(C',i\)$ of $G(C,i)$ induced by $C'$. The edge density of this subgraph is $B'_i$. Since any subset of a code with distance $d$ is by itself a code with distance (at least) $d$, the edge density of $G\(C',i\)$ is upperbounded by $2^{o(n)} \cdot 2^{H\(\frac12 - \sqrt{\delta(1-\delta)}\) \cdot n} \cdot \frac{{n \choose i}}{2^n}$.  Combined with the above lower bound on the edge density of $G(C,i)$, this implies that $G(C,i)$ is $2^{o(n)}$-balanced, and conditions of Conjecture~\ref{cnj:main-general} are satisfied, completing the proof of the proposition.

\eprf

\section{Proof of Propositions~\ref{pro:LLC-random linear}~and~\ref{pro:LLC-random general}}
\label{sec:LLC-random}

\noi We show that Conjectures~\ref{cnj:main-linear}~and~\ref{cnj:main-general} hold for random codes. (See Section~\ref{subsubsec:random} for the models of random codes we use.) In fact they hold in a strong sense, and we are allowed to replace the exponential expressions on the RHS of (\ref{linear}) and (\ref{general}) by absolute constants. We do not attempt to compute the best possible values of these constants.

\subsection{Proof of Proposition~\ref{pro:LLC-random linear}}

\noi We start with a technical lemma. 

\lem
\label{lem:linear - few}

\noi Let $0 < R < 1$ be given. There exist positive constants $\e = \e(R)$ and $\alpha = \alpha(R)$ such that for any integer parameters $d, i, t$ satisfying:
\begin{itemize}

\item $(1 -\e) H^{-1}(1-R) \cdot n \le d \le (1 + \e) H^{-1}(1-R) \cdot n$

\item $d \le i \le (1+\e) d$

\item $d \le t \le 2i$

\end{itemize}

\noi holds: Let $x \in \H$ be of weight $t$. Let $D(x)$ be the set of all points $y$ in $\H$ for which $|y| = |x-y| = i$. Then
\[
|D(x)| ~\le~ 2^{(1 - R - \alpha) \cdot n}.
\]
\elem

\prf

\noi It is easy to see that $|D(x)| = {t \choose \frac{t}{2}} {{n-t} \choose {i - \frac{t}{2}}}$, with the understanding that the binomial coefficient ${b \choose a}$ is $0$ unless $a, b$ are integer and $0 \le a \le b$.

\noi Writing $\delta = \frac dn$, $\xi = \frac in$, and $\tau = \frac tn$, and using (\ref{binom}), it suffices to show that $\tau  + (1-\tau)H\(\frac{2\xi - \tau}{2 - 2\tau}\) < 1-R$ on the compact domain $\Omega = \{(\delta, \xi, \tau) \subseteq \R^3\}$ given by 
\begin{itemize}

\item $(1-\e) H^{-1}(1-R) \le \delta \le (1+\e) H^{-1}(1-R)$

\item $\delta \le \xi \le (1+\e) \delta$

\item $\delta \le \tau \le 2\xi$

\end{itemize}

\noi It is easy to see that for any $0 < R < 1$, if $\e = \e(R)$ is sufficiently small then all the partial derivatives of $f(\xi, \tau) = \tau  + (1-\tau)H\(\frac{2\xi - \tau}{2 - 2\tau}\)$ are uniformly bounded from above on $\Omega$, and hence it suffices to prove that $f(\delta, \tau) < 1-R$ for $\delta \le \tau \le 2\delta$, where $0 < R < 1$ and $\delta = H^{-1}(1-R)$. Alternatively, it suffices to prove that the function $g(\delta, \tau) = \tau + (1-\tau) H\(\frac{2\delta - \tau}{2 - 2\tau}\) - H(\delta)$ is non-positive on $\{(\delta,\tau):~0 \le \delta \le \frac12;~ \delta \le \tau \le 2\delta\}$, and that $g(\delta, \tau) = 0$ only if $\delta = 0$ or $\delta = \frac12$.

\noi We will do this in two steps. First, we claim that $g(\delta,\tau)$ does not increase in $\tau$, for any value of $\delta$. We have, after simplifying, that
\[
\frac{\partial g}{\partial \tau} ~=~ 1 - H\(\frac{2\delta - \tau}{2 - 2\tau}\) - \frac{1-2\delta}{2-2\tau} \log_2\(\frac{2-2\delta-\tau}{2\delta-\tau}\).
\]

\noi Let $x = \frac{2\delta-\tau}{2-2\tau}$. Then the above expression is $1 - H(x) - \(\frac12 - x\)\log_2\(\frac{1-x}{x}\)$. It is easy to see that $H(x) + \(\frac12 - x\)\log_2\(\frac{1-x}{x}\) = \frac12 \log_2\(\frac{1}{x(1-x)}\) \ge 1$, and hence $\frac{\partial g}{\partial \tau} \le 0$.

\noi So it suffices to show that $h(\delta) = g(\delta, \delta)$ is non-positive on $0 \le \delta \le \frac12$, and that $h(\delta) = 0$ only if $\delta = 0$ or $\delta = \frac12$. We have 
\[
h(\delta) ~=~ \delta + (1-\delta) H\(\frac{\delta}{2-2\delta}\) - H(\delta) ~=~
\]
\[
1 + \frac{\delta}{2} \log_2(\delta) + (2-2\delta) \log_2(1-\delta) - \frac{2-3\delta}{2} \log_2(2-3\delta).
\]

\noi It is easy to see that $h(0) = h\(\frac12\) = 0$. We will show that there exists $0 < \delta_0 < \frac12$ such that $h' < 0$ for $0 \le \delta < \delta_0$ and $h' > 0$ for $\delta_0 < \delta \le \frac12$. This will imply that $h(\delta) < 0$ for any $0 < \delta < \frac12$.

\noi A simple calculation gives 
\[
h'(\delta) ~=~ \frac12 \log_2\(\frac{\delta(2-3\delta)^3}{(1-\delta)^4}\).
\]

\noi Let $P(\delta) = \delta(2-3\delta)^3 - (1-\delta)^4$. We need to show that there exists $0 < \delta_0 < \frac12$ such that $P(\delta) < 0$ for $0 \le \delta < \delta_0$ and $P(\delta) > 0$ for $\delta_0 < \delta \le \frac12$.
It is easy to verify that $P(\delta) = (1-2\delta)\(14 \delta^3 - 22 \delta^2 + 10 \delta - 1\)$. So it suffices to show this property for $Q(\delta) = 14 \delta^3 - 22 \delta^2 + 10 \delta - 1$. The derivative $Q'$ is a quadratic, and it is easy to see that it is strictly decreasing on $\left[0,\frac12\right]$, and moreover that $Q'(0) > 0$ and that $Q'\(\frac12\) < 0$. This means that $Q$ is unimodal - it increases up to some point in $\left[0,\frac12\right]$ and then decreases. Moreover, $Q(0) = -1 < 0$ and $Q\(\frac12\) = \frac14 > 0$. This verifies the required property for $Q$, and completes the proof of the lemma.
\eprf

\noi We proceed with the proof of the proposition. Let $f$ be a function on $\H$, and let $A$ be the Fourier support of $f$. By Proposition~1.1 in \cite{ks1} we have $\(\frac{\|f\|_4}{\|f\|_2}\)^4 \le \max_{x \in A+A} \Big |\{(y,z) \in A \times A,\,y+z = x\} \Big |$. So it suffices to show that there exists a positive constant $K = K(R)$ such that the following holds with probability tending to $1$ with $n$ for a random linear code $C$ of dimension $\lfloor Rn \rfloor$.  Let $d$ be the minimal distance of $C$, let $d \le i \le \(1+\e\) d$, where $\e = \e(R)$ is the constant from Lemma~\ref{lem:linear - few}, and let $A$ be the set of vectors of weight $i$ in $C$. Assume $A \not = \emptyset$. Let $f$ be a function whose Fourier transform is supported on $A$. Then
\[
\max_{x \in A+A} \Big |\{(y,z) \in A \times A,\,y+z = x\} \Big | ~\le~ K.
\]

\noi Recall that w.h.p. the minimal distance $d$ of $C$ satisfies $d_0 - o(n)  \le d \le d_0 + o(n)$, where $d_0 = d_0(n) = H^{-1}(1-R) \cdot n$. Assume from now on that this is indeed the case. This means that the points $x$ we need to consider are such that $x = y + z$ for some points $y, z \in \H$ with $|y| = |z| = i$, for some $d_0 -o(n) \le i \le (1+\e) d_0$. Using the union bound, it suffices to show for any suitable value of $i$ and for any such point $x$ the probability over $C$ that $\Big |\{(y,z) \in C \times C,\,|y| = |z| =i,\,y+z = x\} \Big | > K$, for a sufficiently large constant $K$ is $o\(\frac{1}{n2^n}\)$.

\noi Fix $i$ and $x$. Let $|x| = t$. As in Lemma~\ref{lem:linear - few}, let $D(x)$ be the set of all points $y$ in $\H$ for which $|y| = |x-y| = i$. Note that $\Big |\{(y,z) \in C \times C,\,|y| = |z| =i,\,y+z = x\} \Big | \le |D(x) \cap C|$. So it suffices to upperbound $|D(x) \cap C|$.
We proceed to do this. For a subset $D \subseteq \H$ and for an integer parameter $m$, if $|D \cap C| \ge 2^m$, then $C$ contains at least $m$ linearly independent elements of $D$. It is well-known (and easy to see) that the probability of a random linear code $C$ of a given cardinality to contain $m$ given linearly independent vectors is at most $\(\frac{|C|}{2^n}\)^m$, and hence, by the union bound,
\[
\mathrm{Pr}_C\left\{|D \cap C| \ge 2^m\right\} ~\le~ {{|D|} \choose m} \(\frac{|C|}{2^n}\)^m ~<~ \(\frac{|C||D|}{2^n}\)^m.
\]

\noi Let now $D = D(x)$. The parameters $d,i,t$ satisfy the conditions of Lemma~\ref{lem:f-technical}, and hence, by the lemma, $\frac{|C||D|}{2^n} \le 2^{-\alpha n}$, for some positive constant $\alpha = \alpha(R)$. Hence, for $m = \lceil 2/\alpha \rceil$, we get 
\[
\mathrm{Pr}_C\left\{|D \cap C| \ge 2^m\right\} ~\le~ 2^{-\alpha mn} ~\le~ 2^{-2n} ~\le~ o\(\frac{1}{n2^n}\),
\]
as needed. To conclude, the proposition holds with $K = 2^{\lceil 2/\alpha \rceil}$ and $\e_0 = \e$, where $\alpha$ and $\e$ are given by Lemma~\ref{lem:linear - few}.

\eprf

\subsection{Proof of Proposition~\ref{pro:LLC-random general}}

\noi Let $0 < R < 1$. In the following two lemmas we list some simple properties of random codes of prescribed cardinality $N = \lfloor 2^{Rn} \rfloor$ chosen according to the model described in Section~\ref{subsubsec:random}. (We believe all of these properties to be well-known, but we haven't been able to find a proper reference in the literature.)

\noi Before stating the lemmas, let us make some remarks about the parameters $\tau = \tau(R)$ and $d_0 = d_0(R,n)$ in the model. The  constant $\tau$ is chosen to be sufficiently small so that all the constants appearing in the following statements which depend linearly on $\tau$ will be smaller than $\frac12$ and so that $\theta = \frac{{n \choose {d_0}} N}{2^n}$ is at most $\frac12$ as well. It is easy to see that it is possible to choose $\tau$ appropriately. We also note that $\theta$ is an absolute constant depending on $R$ and that $H^{-1}(1-R) \cdot n - o(n) \le d_0 \le H^{-1}(1-R) \cdot n + o(n)$.

\noi We will show that the following properties hold with probability tending to $1$ with $n$ for a random code $C$ of length $n$ and  prescribed cardinality $N = \lfloor 2^{Rn} \rfloor$. We denote by $d$ the minimal distance of $C$.

\lem
\label{lem:C-prop}

\begin{enumerate}

\item $|C| \ge \(1 - O(\tau) - o(1)\) \cdot N$.

\item For any $d \le  k \le \frac n2$ the number of pairs of points in $C$ at distance $k$ from each other lies  between $\(1 - O(\tau) - o(1)\) \cdot \frac{{N \choose 2} {n \choose k}}{2^n}$ and $(1 + o(1)) \cdot \frac{{N \choose 2} {n \choose k}}{2^n}$.
    
\item $d = d_0$.

\end{enumerate}
\elem

\lem
\label{lem:moments ratio}
Let $d \le k \le \frac n2$. For $x \in C$, let $D_{x,k} := |\{y \in C,~|x-y| = k \}|$. Then
\[
\frac{\E_x D^2_{x,k}}{\(\E_x D_{x,k}\)^2} ~\le~ O\(\frac{1}{\theta}\),
\]
where the expectations are taken with respect to the uniform probability distribution on $C$, and $\theta = \frac{{n \choose {d_0}} N}{2^n}$.
\elem
 
\noi We will first prove the proposition assumung Lemmas~\ref{lem:C-prop}~and~\ref{lem:moments ratio} to hold, and then prove the lemmas. 
From now on let $\e = \e(R)$ be the constant in Lemma~\ref{lem:linear - few}.

\ignore{Let $d_0 \le i \le (1+\e)d_0$. We will show that for a sufficiently large constant $K = K(R)$ the probability over $C$ that the vector $\l$ of eigenvalues of the distance-$i$ graph $G(C,i)$ satisfies $\frac{\|\l\|_4}{\|\l\|_2} \le K$ is at least $1 - o\(\frac 1n\)$, which will imply, by the union bound, that the proposition holds with this value of $K$ and with $\e_0 = \e$.
}

\noi We will need another technical lemma. 

\lem
\label{lem:C-mqx-match}
Let $d_0 \le i \le (1+\e) d_0$. For $x, y \in \H$, let $M_{x,y} = |\{z \in C,~|x-z| = |y-z| = i\}|$. Then with probability at least $1 - o\(\frac 1n\)$ we have
\[
\max_{x,y \in C} M_{x,y} ~\le~ K_1,
\]
for some absolute constant $K_1 = K_1(R)$.
\elem

\prf

\noi Let $\tilde{C} = \{x_1,...,x_N\}$ be a list of $N$ points chosen independently at random from $\H$. Clearly, it suffices to prove that there exists a constant $K_1$ such that with high probability for any $\{i,j\} \subseteq [N]$ holds $|M_{x_i, x_j}| \le K_1$. By symmetry, and by the union bound argument, it suffices to show that $\mathrm{Pr}_{\tilde{C}}\left\{|M_{x_1, x_2}| > K_1\right\} < \frac{1}{n N^2}$, where we may assume that $|x_1 - x_2| \ge d_0$. Fix $x_1$ and $x_2$. Let $|x_1 - x_2| = t$ for some $d_0 \le t \le 2i$ and let $D = D(x,y)$ be the set of all points in $\H$ at distance $i$ from both $x$ and $y$. Then $|D| = {t \choose \frac{t}{2}} {{n-t} \choose {i - \frac{t}{2}}}$. Let $\tilde{C}_1 = \{x_3,...,x_N\}$. For any integer $m \ge 1$ we have that
\[
\mathrm{Pr}_{\tilde{C}}\left\{|M_{x,y}| \ge m\right\} ~=~ \mathrm{Pr}_{\tilde{C}_1}\left\{|\tilde{C}_1 \cap D| \ge m\right\} ~\le~ {{|D|} \choose m} \cdot \(\frac{|\tilde{C}_1|}{2^n}\)^m ~<~ \(\frac{ND}{2^n}\)^m.
\]

\noi By Lemma~\ref{lem:linear - few} $\frac{ND}{2^n} \le 2^{-\alpha n}$, for a positive absolute constant $\alpha$, and hence for $K_1 = \frac{3}{\alpha}$, we have $\mathrm{Pr}\left\{|M_{x,y}| > K_1\right\} < 2^{-3n} < \frac{1}{n N^2}$, and the claim of the lemma holds with this value of $K_1$. 
\eprf

\noi We proceed to prove the proposition. Let $d_0 \le i \le (1+\e) d_0$, and let $B$ be the adjacency matrix of the graph $G(C,i)$. Let $\l$ be the vector of eigenvalues of $B$. Recall the notation $D_{x,i} := |\{y \in C,~|x-y| = i\}|$ and $M_{x,y} = |\{z \in C,~|x-z| = |y-z| = i\}|$. We have
\[
\|\l\|^2_2 ~=~ \frac{1}{|C|} \sum_{i=1}^{|C|} \l^2_i ~=~ \frac{1}{|C|} \sum_{x,y \in C} B^2(x,y) ~=~ \frac{1}{|C|} |\{(x,y) \in C \times C,\, |x-y| = i\}| ~=~ \E_{x \in C} D_{x,i}.
\]

\noi For the following calculation, let $g(z) = 1_{|z| = i}$. Then for any $x,y \in C$ we have $B(x,y) = g(x+y)$. Hence, using the Cauchy-Schwarz inequality in the fourth step below,
\[
\|\l\|^4_4 ~=~ \frac{1}{|C|} \sum_{i=1}^{|C|} \l^4_i ~=~ \frac{1}{|C|} \sum_{x,y \in C} \(\sum_{z \in C} B(x,z)B(z,y)\)^2 ~=~
\]
\[
\frac{1}{|C|} \sum_{x,y \in C} \(\sum_{z \in C} g(x+z)g(y+z)\)^2 ~\le~ 
\frac{1}{|C|}  \sum_{x,y \in C} M_{x,y} \sum_{z \in C} g^2(x+z) g^2(y+z) ~\le~
\]
\[
\frac{\max_{x,y \in C} M_{x,y}}{|C|} \cdot \sum_{x,y \in C} \sum_{z \in C} g^2(x+z) g^2(y+z) ~=
\]
\[
\frac{\max_{x,y \in C} M_{x,y}}{|C|} \cdot \sum_{z \in C} \(\sum_{x \in C} g^2(x+z)\)^2 ~=~ \(\max_{x,y \in C} M_{x,y}\) \cdot \E_{z \in C} D^2_{z,i}.
\]

\noi It follows that
\[
\(\frac{\|\l\|_4}{\|\l\|_2}\)^4 ~\le~ \(\max_{x,y \in C} M_{x,y}\) \cdot \frac{\E_z D^2_x}{\(\E_z D_z\)^2}.
\]

\noi Lemmas~\ref{lem:moments ratio}~and~\ref{lem:C-mqx-match} imply that with probability tending to $1$ with $n$ the RHS of the last inequality is bounded from above by $O\(\frac{1}{\theta} \cdot K_1\)$ for all $d_0 \le i \le (1+\e) d_0$, and the proposition holds with $K = O\(\frac{1}{\theta} \cdot K_1\)$ and $\e_0 = \e$.

\eprf

\noi We proceed with the proofs of Lemmas~\ref{lem:C-prop}~and~\ref{lem:moments ratio}.

\subsubsection{Proof of Lemmas~\ref{lem:C-prop}~and~\ref{lem:moments ratio}}

We need the following technical claim. 
 
\lem
\label{lem:technical-dist-distr}
Let $\tilde{C} = \{x_1,...,x_N\}$ be a list of $N$ points chosen independently at random from $\H$. Let $M = {N \choose 2}$, and for $0 \le \ell \le n$ let $p_{\ell} = \frac{{n \choose {\ell}}}{2^n}$.

\begin{enumerate}

\item For $0 \le \ell \le n$ let $X_{\ell}$ be the random variable counting the number of pairs of indices $1 \le i < j \le N$ with $|x_i - x_j| = \ell$. Then $\E X_{\ell} = M p_{\ell}$, and $\sigma^2\(X_{\ell}\) = M p_{\ell}\(1-p_{\ell}\)$.

\item For $d_0 \le \ell \le \frac n2$, let $Y_{\ell}$ be the random variable counting the number of triples of distinct indices $\{a,b,c\} \subseteq [N]$ such that one of the three pairwise distances among the points $x_a, x_b, x_c$ is $\ell$ and one is at most $d_0-1$. Then $\E Y_{\ell} \le O\(\tau \cdot Mp_{\ell}\)$, and $\sigma^2\(Y_{\ell}\) \le O\(\tau M p_{\ell} + \tau M^{3/2} p_{\ell}^2\)$.

\end{enumerate}
\elem

\prf

\noi For $\{i,j\} \subseteq N$ let $Z_{\{i,j\}}$ be the indicator random variable which verifies whether $|x_i - x_j| = \ell$. Clearly $Pr_{\tilde{C}} \left\{Z_{\{i,j\}} = 1\right\} = \frac{{n \choose {\ell}}}{2^n} = p_{\ell}$. The first claim of the lemma follows from the fact that there are $M$ such variables, and (as observed e.g., in \cite{bf}) they are pairwise independent.

\noi We pass to the second claim of the lemma. For $\{a,b,c\} \subseteq N$ let $Z_{\{a,b,c\}}$ be the indicator random variable which verifies whether one of the three pairwise distances among the points $x_a, x_b, x_c$ is $\ell$ and one is at most $d_0-1$. We have, by the pairwise independence of the distances $|x_a - x_b|$, $|x_a - x_c|$, and $|x_b - x_c|$, that 
\[
\E_{\tilde{C}} Y_{\ell} ~=~ \sum_{\{a,b,c\} \subseteq N} \E_{\tilde{C}} Z_{\{a,b,c\}} ~\le~ O\(N^3 \cdot \frac{\(\sum_{k=0}^{d_0-1} {n \choose k}\) \cdot  {n \choose {\ell}}}{2^{2n}}\) ~\le~ 
\]
\[
O\(M p_{\ell} \cdot N \frac{\sum_{k=0}^{d_0-1} {n \choose k} }{2^n}\) ~\le~ O\(\tau \cdot Mp_{\ell}\),
\]
where the last step follows from the choice of $d_0$.

\noi Next, observe that for $\{a,b,c\}, \{a',b',c'\} \subseteq N$ the random variables $Z_{\{a,b,c\}}$ and $Z_{\{a',b',c'\}}$ are independent unless $|\{a,b,c\} \cap \{a',b',c'\}| \ge 2$. Hence
\[
\sigma^2\(Y_{\ell}\) ~=~ \E Y^2_{\ell} - \(\E Y_{\ell}\)^2 ~=~ \sum_{\{a,b,c\}, \{a',b',c'\}} \Big(\E Z_{\{a,b,c\}} \cdot Z_{\{a',b',c'\}} - \E Z_{\{a,b,c\}} \cdot \E Z_{\{a',b',c'\}}\Big) ~=~
\]
\[
\sum_{|\{a,b,c\} \cap \{a',b',c'\}| \ge 2} \Big(\E Z_{\{a,b,c\}} \cdot Z_{\{a',b',c'\}} - \E Z_{\{a,b,c\}} \cdot \E Z_{\{a',b',c'\}}\Big) ~\le
\]
\[
\sum_{|\{a,b,c\} \cap \{a',b',c'\}| \ge 2} E Z_{\{a,b,c\}} \cdot Z_{\{a',b',c'\}} ~=~ \sum_{\{a,b,c\}} \E_{\tilde{C}} Z_{\{a,b,c\}} + \sum_{|\{a,b,c\} \cap \{a',b',c'\}| = 2} E Z_{\{a,b,c\}} \cdot Z_{\{a',b',c'\}}.
\]

\noi The first summand in the last expression is at most $O\(\tau \cdot Mp_{\ell}\)$. We pass to the second summand. There are two possible cases we need to consider, depending on whether $\sum_{k=0}^{d_0-1} {n \choose k}$ is smaller than ${n \choose {\ell}}$. Assume first that it is indeed smaller. For $\{a,b,c,d\} \subseteq N$ let $W_{\{a,b,c,d\}}$ be the indicator random variable which verifies whether there are two among the four points at distance at most $d_0-1$, and each of the two remaining points is at distance $\ell$ from one of the first two. It is easy to see that the second summand is upperbounded by
\[
O\(\sum_{\{a,b,c,d\} \subseteq N} \E W_{a,b,c,d}\) ~\le~ O\(N^4 p_{\ell}^2 \frac{\sum_{k=0}^{d_0-1} {n \choose k}}{2^n}\) ~\le~ O\(\tau N^3 p_{\ell}^2\) ~=~ O\(\tau M^{3/2} p_{\ell}^2\).
\]

\noi If $\sum_{k=0}^{d_0-1} {n \choose k}$ is larger than ${n \choose {\ell}}$, a similar computation shows that the second summand is upperbounded by $O\(\tau^2 \cdot Mp_{\ell}\)$, completing the proof of the lemma. 
\eprf

\noi We can now prove Lemma~\ref{lem:C-prop}.

\prf (Lemma~\ref{lem:C-prop})

\noi The first claim of the lemma follows by the Chebyshev inequality from the first claim of Lemma~\ref{lem:technical-dist-distr} (and the definition of $d_0$). The third claim of the lemma follows from its second claim. We pass to the second claim. Fix $d \le k \le \frac n2$. Observe first that by the first claim of Lemma~\ref{lem:technical-dist-distr} and the Chebyshev inequality, the number of pairs of points in $\tilde{C}$ at distance $k$ from each other lies  between $(1 - o(1)) \cdot Mp_k$ and $(1 + o(1)) \cdot Mp_k$ with probability at least $1 - o\(\frac 1n\)$. Next, note that the number of pairs of points in $\tilde{C}$ at distance $k$ from each other removed in the erasure step is at most $O\(Y_k\)$. By the second claim of Lemma~\ref{lem:technical-dist-distr} and the Chebyshev inequality, $Y_k \le  O\(\tau M p_k\)$ with probability at least $1 - o\(\frac 1n\)$, and the second claim of the lemma follows, by the union bound over all possible values of $k$.

\eprf

\noi We pass to the proof of Lemma~\ref{lem:moments ratio}.

\prf (Lemma~\ref{lem:moments ratio})

\noi Fix $d \le k \le \frac n2$. We will show that $\mathrm{Pr}_C \left\{\frac{\E_x D^2_{x,k}}{\(E_x D_{x,k}\)^2} ~\le~ O\(\frac{1}{\theta}\)\right\} \ge 1 - o\(\frac 1n\)$, and the claim of the lemma will follow by the union bound over all possible values of $k$. For notational convenience we will write $D_x$ for $D_{x,k}$ in the remainder of the proof, and we will write 'with high probability' (w.h.p.) for probability at least $1 - o\(\frac 1n\)$.

\noi First, we have that $\sum_{x \in C} D_x = \sum_{x \in C} |\{y \in C,~|x+y| = k \}|$ is the number of pairs of points in $C$ at distance $k$ from each other. As observed in the proof of the second claim of Lemma~\ref{lem:C-prop}, this number is w.h.p. at least $\Omega\(M p_k\)$. This implies that $\E_x D_x = \frac{1}{|C|} \sum_{x \in C} D_x \ge \Omega\(N p_k\)$, and hence $\(\E_x D_x\)^2 \ge \Omega\(N^2 p^2_k\)$. We will show that, w.h.p. $\E_x D^2_x \le O\(\frac{1}{\theta} \cdot N^2 p^2_k\)$, and this will imply the claim of the lemma.

\noi Let $\tilde{C} = \{x_1,...,x_N\}$ be a list of $N$ points chosen independently at random from $\H$. For $i \in [N]$, let $\tilde{D}_i$ be the number of indices $j \in [N]$ so that $|x_i - x_j| = k$. It suffices to show that w.h.p. $\sum_{i=1}^N \tilde{D}^2_i \le  O\(\frac{1}{\theta} \cdot N^3 p^2_k\)$. We proceed similarly to the proof of Lemma~\ref{lem:technical-dist-distr}. Let $S = \sum_{i=1}^N \tilde{D}^2_i$. For $\{i,j\} \subseteq N$ let $Z_{\{i,j\}}$ be the indicator random variable which verifies whether $|x_i - x_j| = k$. We have that
\[
\E_{\tilde{C}} S ~=~ \E_{\tilde{C}} \sum_{i=1}^N \(\sum_{j=1}^N Z_{i,j}\)^2 ~=~ \sum_{i=1}^N \sum_{j_1, j_2 =1}^N \E_{\tilde{C}} \(Z_{i,j_1} \cdot Z_{i,j_2}\) ~=~
\]
\[
\sum_{i=1}^N \sum_{j=1}^N \E_{\tilde{C}} Z_{i,j} + \sum_{i=1}^N \sum_{j_1 \not = j_2} \E_{\tilde{C}} Z_{i,j_1} \cdot E_{\tilde{C}} Z_{i,j_2} ~\le~ O\(N^2 p_k\) + O\(N^3 p^2_k\).
\]

\noi Note that since $d_0 \le k \le \frac n2$, we have that $N p_k \ge N p_{d_0} \ge \theta$ and hence the bound above is at most $O\(\frac{1}{\theta} \cdot N^3 p^2_k\)$.

\noi Next, we claim that $\sigma^2(S) \le O\(N^4 p_k^3\)$. The argument for this estimate is very similar to that for the bound on the variance of $Y_{\ell}$ in the proof of the second claim of Lemma~\ref{lem:technical-dist-distr} and we omit it. This bound on the variance implies, via Chebyshev's inequality, that w.h.p. $S \le O\(\frac{1}{\theta} \cdot N^3 p^2_k\)$, completing the proof of the lemma.

\eprf

\subsubsection*{Acknowledgement}

\noi We are grateful to Johan H{\aa}stad for his permission to publish his conjecture (see (\ref{hastad})) in our paper.


\begin{thebibliography}{99}



\bibitem{abl}
A. Ashikhmin, A. Barg, and S. Litsyn, {\sl Estimates of the distance distribution
of codes and designs}, IEEE Trans. Inf. Theory, vol. 47, no. 2, pp. 1050-1061, Mar. 2001.

\bibitem{abv}
A. Ashikhmin, A. Barg, and S. G. Vladuts,
{\sl Linear codes with many light vectors}, Journal of Combinatorial
Theory, Ser. A, vol. 96, 2, 2001.

\bibitem{bgsv}
C. Bachoc, D. C. Gijswijt, A. Schrijver, and F. Vallentin, {\sl Invariant semidefinite programs}, in  Anjos M., Lasserre J. (eds) {\bf Handbook on Semidefinite, Conic and Polynomial Optimization}, International Series in Operations Research and Management Science, vol 166. Springer, Boston, MA.

\bibitem{bf}
A. Barg and G. D. Forney, {\sl Random Codes: Minimum Distances and Error Exponents}, IEEE Trans. on Inform. Th., 48(9), 2002.

\bibitem{bj}
A. Barg and D. B. Jaffe, {\sl Numerical results on the asymptotic rate of binary codes},
in ``Codes and Association Schemes'' (A. Barg and S. Litsyn, Eds.), Amer. Math. Soc.,
Providence, 2001.

\bibitem{biggs}
N. Biggs, {\bf Algebraic graph theory}, Cambridge University Press, 1974.

\bibitem{dels}
P. Delsarte, {\sl An algebraic approach to the association schemes of
coding theory}, Philips Res. Rep., Suppl., vol. 10, 1973.

\bibitem{O'Donnel}
R. O'Donnel, {\bf Analysis of Boolean functions}, Cambridge University Press, 2014.

\bibitem{ft}
J. Friedman and J-P. Tillich, {\sl Generalized Alon-Boppana theorems and
error-correcting codes}, SIAM J. Discrete Math., 19(3) (electronic), 2005, pp. 700-718.

\bibitem{gms}
D. C. Gijswijt, H. D. Mittelmann, and A. Schrijver,{\sl Semidefinite code bounds based on quadruple
distances}, IEEE Trans. on Inform. Th., 58(5), 2697 - 2705.


\bibitem{Hastad-personal}
J. Hastad, personal communucation, 2016.

\bibitem{it}
P. Ivanisvili and T. Tkocz, {\sl Comparison of moments of Rademacher Chaoses},
arXiv preprint arXiv:1807.04358, 2018

\bibitem{kal:lin}
G. Kalai and N. Linial, {\sl On the distance distribution of codes},
IEEE Trans. Inform. Theory, vol. IT-41, 1995, 1467-1472.

\bibitem{khintchine}
A. Khintchine: {\sl Uber dyadische Br\"uche}, Math. Z. 18, 109-116 (1923)

\bibitem{ks1}
N. Kirshner and A. Samorodnitsky, {\sl On $\ell_4 : \ell_2$ ratio of functions with restricted Fourier support},  J. Comb. Theory, Ser. A 172, pp. 105-202 (2020)

\bibitem{ks2}
N. Kirshner and A. Samorodnitsky, {\sl A moment ratio bound for polynomials and some extremal
properties of Krawchouk polynomials and Hamming spheres}, IEEE Trans. Inform. Theory to appear.

\bibitem{lev-chapter}
V. I. Levenshtein, {\sl Universal bounds for codes and designs}, in ``Handbook of Coding
Theory'' (V. S. Pless and W. C. Huffman, Eds.), Elsevier, Amsterdam, 1998.

\bibitem{ls}
N. Linial and A. Samorodnitsky, {\sl Linear codes and sums of characters}, Combinatorica 22(4), 2002, 497-522.

\bibitem{van lint}
J.H. van Lint, {\bf Introduction to Coding Theory}, third edition, Graduate Texts in Mathematics,
vol. 86, Springer-Verlag, Berlin, 1999.

\bibitem{macw-sl}
J. MacWilliams and N. J. A. Sloane, {\bf The Theory of Error Correcting Codes}, Amsterdam, North-Holland, 1977.

\bibitem{mrrw}
R. J. McEliece, E. R. Rodemich, H. Rumsey, Jr., and L. R. Welch,
{\sl New upper bounds on the rate of a code via the Delsarte-MacWilliams
inequalities}, IEEE Trans. Inform. Theory, vol. 23, 1977, pp. 157-166.

\bibitem{ns}
M. Navon, A. Samorodnitsky, {\sl On Delsarte's linear programming bounds for binary codes}, Proceedings of FOCS 46.

\bibitem{ns1}
M. Navon, A. Samorodnitsky, {\sl Linear programming bounds for codes via a covering argument}, Disc. and Comp. Geom. 41(2): 199-207 (2009).

\bibitem{sam-log-sob}
A. Samorodnitsky, {\sl A modified logarithmic Sobolev inequality for the Hamming cube and some applications}, preprint arXiv:0807.1679 (2008).

\bibitem{schrijver}
A. Schrijver, {\sl New code upper bounds from the Terwilliger algebra
and semidefinite programming}, IEEE Trans. on Inform. Th., 51, 2005, pp. 2859-2866.

\bibitem{t}
C. M. Theobald, {\sl An inequality for the trace of the product of two symmetric matrices}, Math. Proceedings Cambridge Phil. Soc., vol.  77, 02, pp. 265 - 267, 1975.

\end{thebibliography}
\end{document}